\documentclass[12pt,twoside,pointlessnumbers,nofootinbib,smallheadings]{revtex4}
\usepackage{amsmath}
\usepackage{amssymb}
\usepackage{setspace}
\usepackage{hyperref}

\usepackage{hangcaption,fancybox,feynmp}
\usepackage{indentfirst}
\usepackage{graphicx}
\usepackage{epsfig,subfigure,psfrag}
\usepackage{mathrsfs}

\begin{document}

\title{
A simple algorithm for automatic Feynman diagram generation}
\author{Bo Xiao $^{1}$,
Hao Wang $^{2}$\footnote{E-mail:hwang@mail.bnu.edu.cn},
and Shou-hua Zhu$^{3,4}$ }

\affiliation{ $ ^1$ Institute of Fluid Physics, China Academy of Engineering Physics, Mianyang 621900, China \\
$ ^2$ Department of Astronomy, Beijing Normal University, Beijing 100875, China\\
$ ^3$ Institute of Theoretical Physics $\&$ State Key Laboratory of Nuclear Physics and Technology, Peking University, Beijing 100871, P. R. China\\
$ ^4$ Center for High Energy Physics, Peking University, Beijing 100871, P. R. China }

\date{\today}

\maketitle

\begin{center}
{\bf Abstract}

\begin{minipage}{15cm}
{\small  \hskip 0.25cm

An algorithm for the automatic Feynman diagram (FD) generation is presented in this paper. The algorithm starts directly from the definition formula of FD, and is simple in concept and easy for coding. The symmetry factor for each FD is naturally generated. It is expected to bring convenience for the researchers who are studying new calculation techniques or making new calculation tools and for the researchers who are studying effective field theory. A C-program made from the algorithm is also presented, which is short, fast, yet very general purpose: it receives arbitrary user defined model and arbitrary process as input and generates FD's at any order.

}

\end{minipage}
\end{center}


\newpage
\section{Introduction\label{Introduction}}

Cross section calculation is an important and hard task in high energy physics. The calculation technique in high energy physics is based on the Lorentz invariant perturbative expansion of the S-matrix developed by Feynman, Schwinger, and Tomonaga. In this technique, a calculation begins with generating all the FD's. The number of Feynman diagrams for a normal scattering process can be hundreds or thousands. To obtain all of these diagrams manually would be exhausting and prone to errors. The Calculations of these FD's are even more tiring. Thus, many software's have appeared which can manage these steps automatically (see \cite{Harlander:1998dq} for a good review).

The rules for FD generation are definite, hence are quite suitable for compute automation. There are well known general purpose FD generators such as FeynArts \cite{Kublbeck:1990xc,Hahn:2000kx}, QGRAF \cite{Nogueira:1991ex} and the grc part of GRACE \cite{Kaneko:1994fd}. FeynArts and QGRAF generate FD's with the steps of: 1. generating all the connected topologies; 2. eliminating equivalent topologies generated; 3. inserting physical fields into the topologies to generate all the FD's; 4. eliminating equivalent FD's generated. The grc part of GRACE generates FD's by: 1. uses an iteratively connecting method to generate all the FD's; 2. eliminate equivalent FD's generated. In all these three generators, the post eliminating procedure is time consuming and full of tricks. There are also many more not so general FD generators \cite{Boos:1994xb,Pukhov:1999gg,Stelzer:1994ta,Krauss:2001iv,Wang:2004du,Schelstraete:1995tk, Kleinert:1999uv,Kajantie:2001hv}. MadGraph \cite{Stelzer:1994ta} and AMEGIC \cite{Krauss:2001iv} adopt a topology-based algorithm. They are restricted to the tree diagrams. The FD generators in \cite{Schelstraete:1995tk,Kleinert:1999uv,Kajantie:2001hv} are novel and fast, but have limited functionalities and are designed for some special purposes. Overall, the existing algorithms for general purpose FD generation are some complicated that are not easy for other researchers to follow to write their own FD generators.

A concept-simple and coding-easy FD algorithm could prove to be helpful to, say, researchers developing new calculation tools. For example, some researchers have being developing new techniques of doing the loop integration in a numerical way \cite{Kilian:2009wy}; using self-made FD generators would make the process of such studies more convenient.

The rest of the paper is organized as follows. In section \ref{sec.algorithm}, the simple FD generation algorithm is described; In section \ref{sec.Cprogram}, a small C-program that realize this algorithm is described; In section \ref{sec.checkment}, the correctness of the algorithm/C-program is checked; In section \ref{sec.conclusion},  conclusions are given.

\section{the algorithm for Feynman diagram generation\label{sec.algorithm}}

The principles for FD generation is contained in the well-known perturbative formula for the S-matrix \cite{Weinberg:1995}.
\begin{equation}
 \begin{array}{lr}
 {S_{{p^\prime_1}\sigma^\prime_1{n^\prime_1};{p^\prime_2}\sigma^\prime_2{n^\prime_2}; \cdots {p_1}{\sigma _1}{n_1};{p_2}{\sigma _2}{n_2}; \cdots }}\\
 = \sum\limits_{N = 0}^\infty  {\frac{{{{( - i)}^N}}}{{N!}}\int {{d^4}{x_1} \cdots {d^4}{x_N}} } \left( {\Phi _0} \cdots a({p^\prime_2}{\sigma^\prime_2}{n^\prime_2})a({p^\prime_1}{\sigma^\prime_1}{n^\prime_1})\right.\\
 \;\;\;\;\left.\times T\left\{ {H({x_1}) \cdots H({x_N})} \right\}{a^\dag }({p_1}{\sigma_1}{n_1}){a^\dag }({p_2}{\sigma_2}{n_2}) \cdots {\Phi _0} \right).
 \end{array}
 \label{Eq.start}
\end{equation}
It's convenient to convert (\ref{Eq.start}) into an abbreviate form, since one concerns only the FD generation here
\begin{equation}
 S_{1^\prime 2^\prime  \cdots 12 \cdots }^N \sim \frac{1}{N!}\{ \cdots a^\prime_2 a^\prime _1\} T\{H_1 \cdots H_N\} \{a_1^\dag a_2^\dag  \cdots \}
\end{equation}
It is also convenient to turn all the particles in the ``in'' state into its antiparticles in the ``out'' state
\begin{equation}
 S_{12 \cdots n}^N \sim \frac{1}{{N!}}\left\{ {a_n a_{n-1} \cdots {a^c_2}{a^c_1}} \right\} T\left\{ {{H_1} \cdots {H_N}} \right\},
 \label{Eq.abbrSN},
\end{equation}
where the primes have been discarded and subscripts have been modified. This conversion do not change the structure of a FD. Eq. (\ref{Eq.abbrSN}) is the base formula for the following discussion.

Let's consider a FD generation example for the process $e_1\bar{e}_2\rightarrow e_3\bar{e}_4$ and the number of interaction vertex $N=2$ and the model
\begin{equation}
 H \propto {\bar \psi ^e}{\psi ^e}{\phi ^n}
 \label{Eq.H1}
\end{equation}
In this example, the expression (\ref{Eq.abbrSN}) can be put into an explicit form as shown in Eq. (\ref{Eq.explicitSNforH1}) (note: $e_1$ and $\bar{e}_2$ in the ``in'' state have been turned into their antiparticles $\bar{e}_1$ and $e_2$ in the ``out'' state in Eq. (\ref{Eq.explicitSNforH1}))
\begin{equation}
 S_{1234}^2 \sim \frac{1}{{2!}}\left\{ {{{\bar e}_4}{e_3}{e_2}{{\bar e}_1}} \right\} T\left\{ {\left[ {{{\bar \psi }_1}{\psi _1}{\phi _1}} \right] \left[ {{{\bar \psi }_2}{\psi _2}{\phi _2}} \right]} \right\}
 \label{Eq.explicitSNforH1}
\end{equation}
All the required FD's are contained in expression (\ref{Eq.explicitSNforH1}), that is, they correspond to all the combinations of grouping the annihilation operators and the field operators according to the pairing
\[
(\bar e\psi)\text{, }(e\bar\psi)\text{, }(n\phi)\text{, }(\psi\bar\psi)\text{, and }(\phi\phi).
\]

To find all the combinations, we define an order for the pairing procedure

\textbf{Pairing order:}
\begin{itemize}
  \item Firstly, the external particles are scanned one by one from right to left (from the ones with smaller subscript to the ones with larger subscript) to pair with the fields in the T\{\} brace;
  \item Secondly, after all the external particles are scanned, the remaining unpaired fields in the T\{\} brace are scanned one by one from left to right (from smaller subscript to larger subscript) to pair with other fields
  \item When an operator (an external particle or a field in the T\{\} brace) is pairing with the fields in the T\{\} brace, it scans the fields from left to right.
\end{itemize}

\begin{figure}[htbp]
\centering
\subfigure[]{\includegraphics[width=0.2\textwidth]{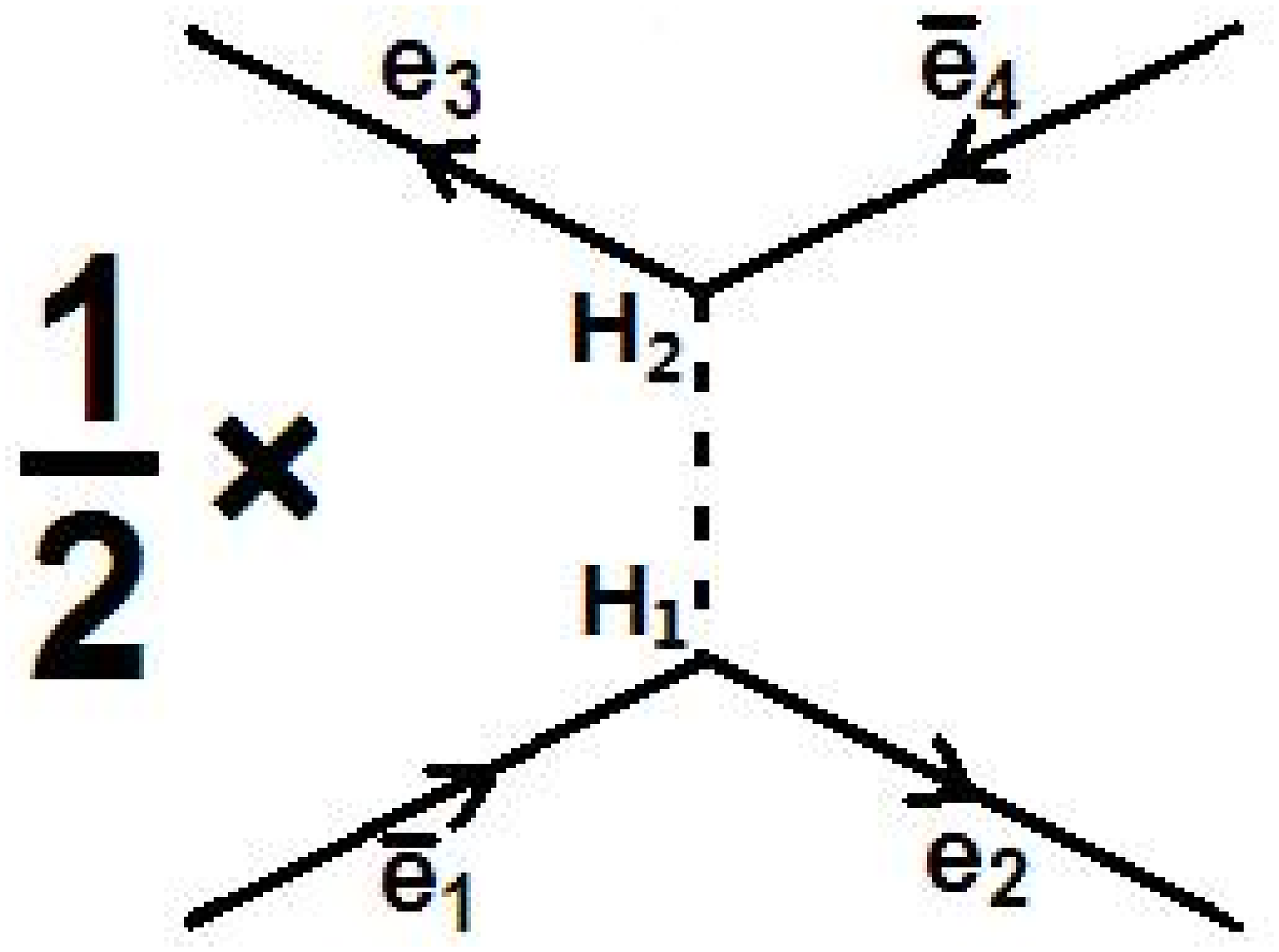}}
\subfigure[]{\includegraphics[width=0.2\textwidth]{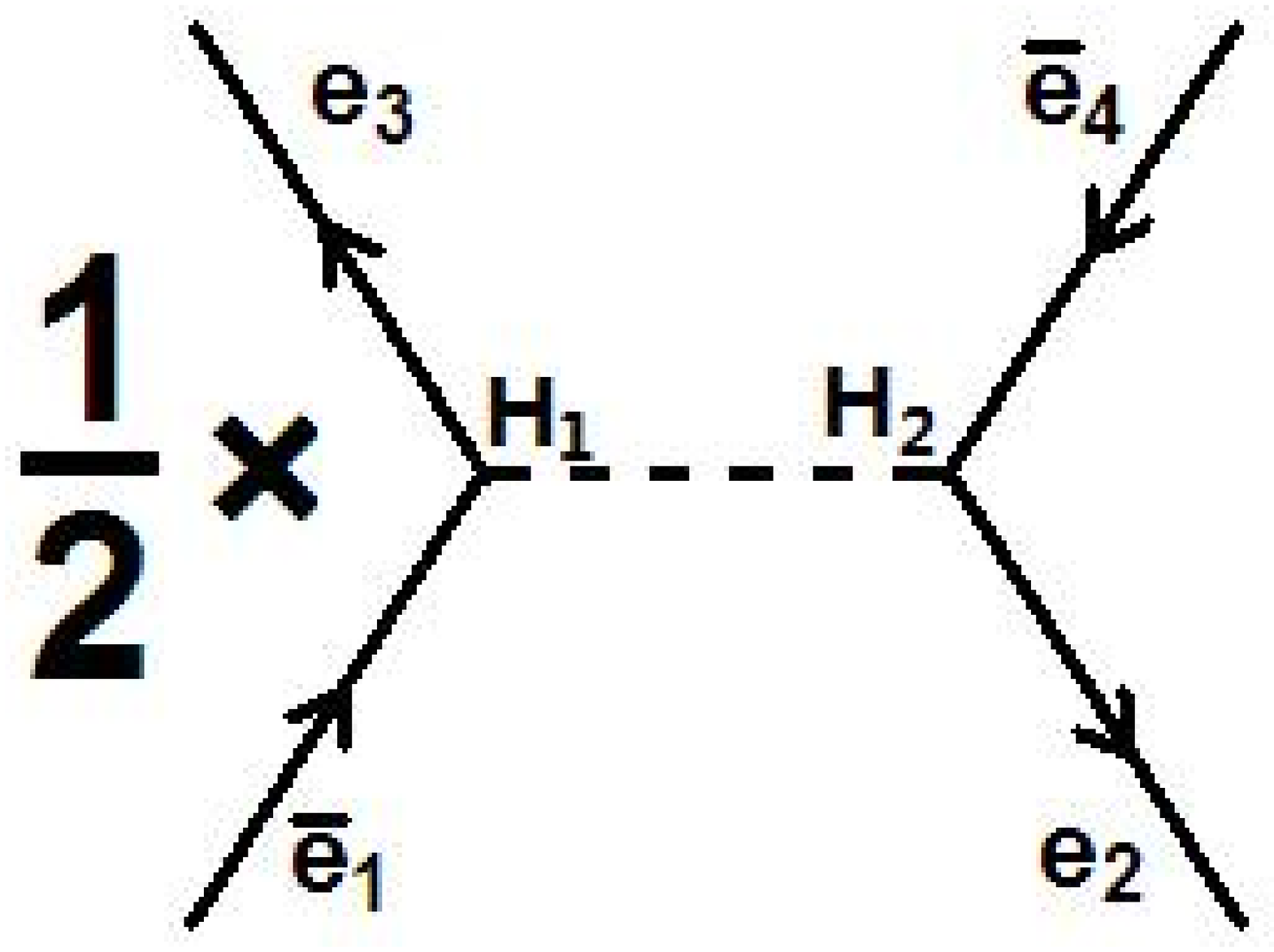}}
\subfigure[]{\includegraphics[width=0.2\textwidth]{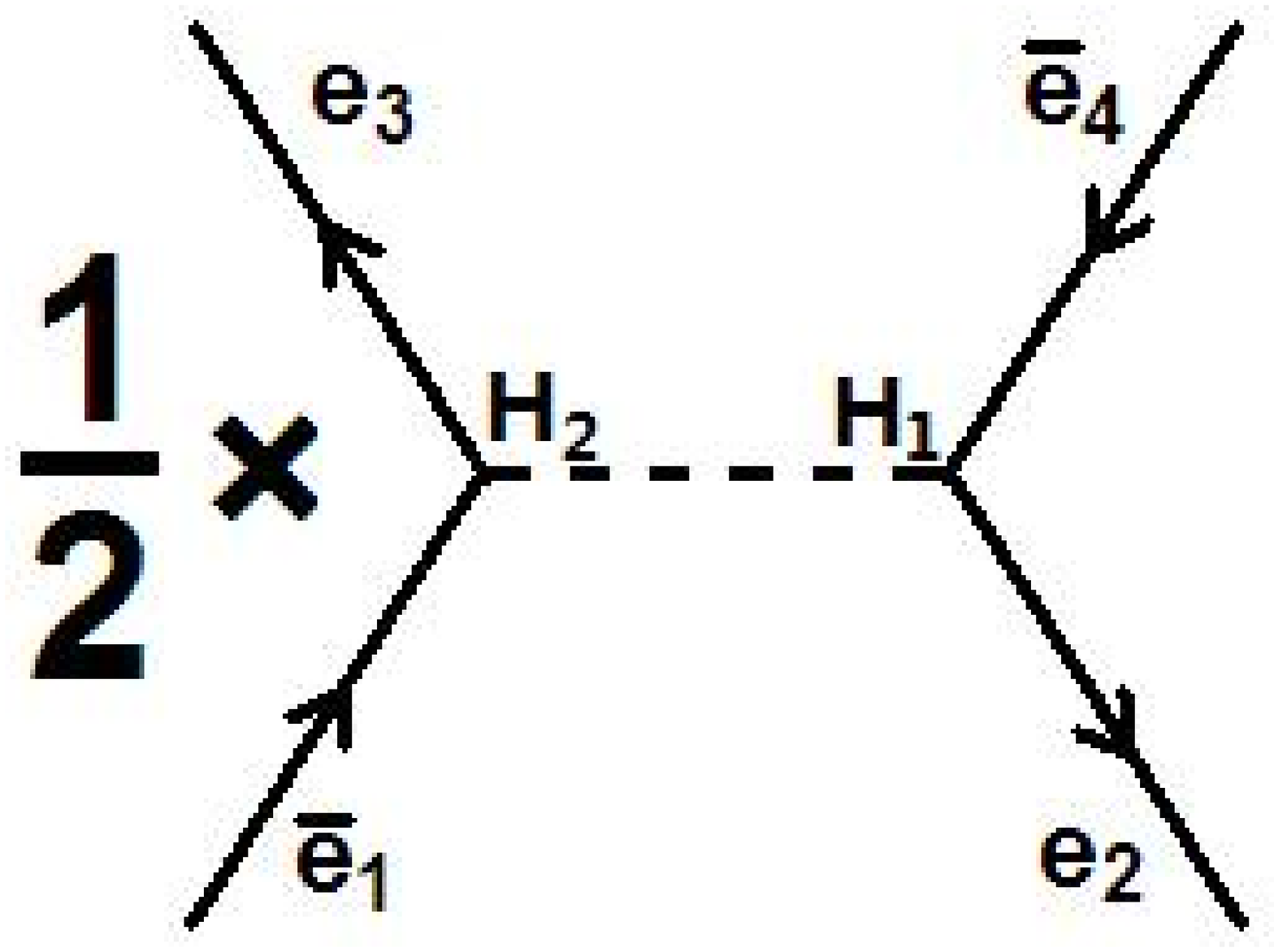}}
\subfigure[]{\includegraphics[width=0.2\textwidth]{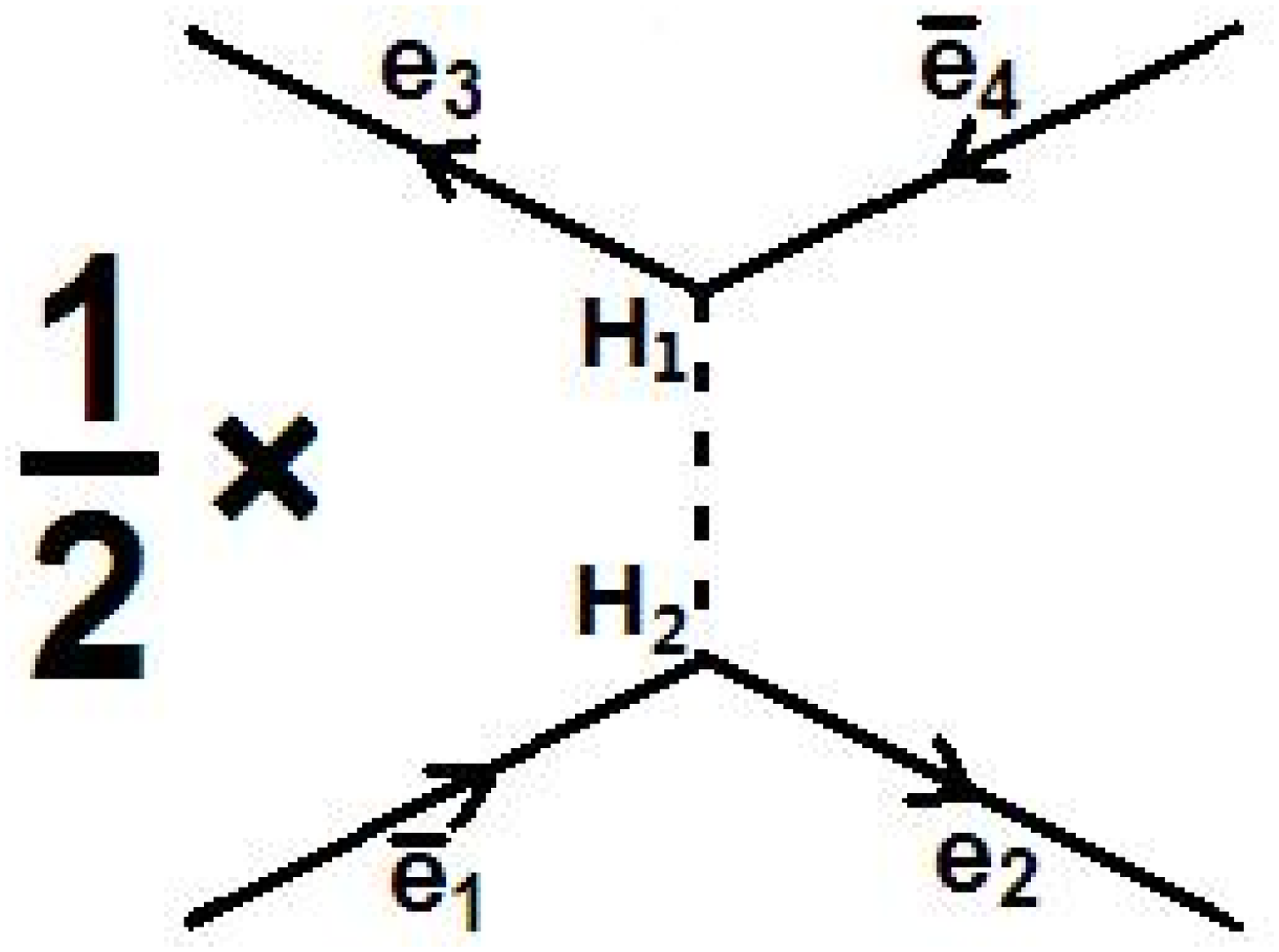}}
\caption{\label{FourDiagrams}four Feynman diagrams corresponding to the four combinations in (6). }
\end{figure}

Following this pairing procedure we got for Eq. (\ref{Eq.explicitSNforH1}) four combinations, they are put in Eq. (\ref{Eq.fourDiag}) in the order of their production
\begin{equation}
 \begin{array}{l}
 \frac{1}{{2!}}T\left\{ {\left[ {({e_2}{{\bar \psi }_1})({{\bar e}_1}{\psi _1})} \right] \left[ {({e_3}{{\bar \psi }_2})({{\bar e}_4}{\psi _2})({\phi _1}{\phi _2})} \right]} \right\}\\
 \frac{1}{{2!}}T\left\{ {\left[ {({e_3}{{\bar \psi }_1})({{\bar e}_1}{\psi _1})} \right] \left[ {({e_2}{{\bar \psi }_2})({{\bar e}_4}{\psi _2})({\phi _1}{\phi _2})} \right]} \right\}\\
 \frac{1}{{2!}}T\left\{ {\left[ {({e_2}{{\bar \psi }_1})({{\bar e}_4}{\psi _1})} \right] \left[ {({e_3}{{\bar \psi }_2})({{\bar e}_1}{\psi _2})({\phi _1}{\phi _2})} \right]} \right\}\\
 \frac{1}{{2!}}T\left\{ {\left[ {({e_3}{{\bar \psi }_1})({{\bar e}_4}{\psi _1})} \right] \left[ {({e_2}{{\bar \psi }_2})({{\bar e}_1}{\psi _2})({\phi _1}{\phi _2})} \right]} \right\}
 \end{array}
 \label{Eq.fourDiag}
\end{equation}
The four combinations in Eq. (\ref{Eq.fourDiag}) correspond to four diagrams in Figure \ref{FourDiagrams} respectively.

In Figure \ref{FourDiagrams} we see, (a) and (d) (or (b) and (c)) differ only in the vertices labels, so they are topologically equivalent diagrams. These equivalent diagrams can be avoided easily by a simple trick. Let's look again with more care at the pairing procedure that leads Eq. (\ref{Eq.explicitSNforH1}) to Eq. (\ref{Eq.fourDiag}). The external particle $\bar{e}_1$ is to be paired with a field $\psi$, and it faces two choices: pairing with $\psi_1$ in $H_1$ or with $\psi_2$ in $H_2$, that is
\begin{equation}
 \begin{array}{l}
 \left\{ {{{\bar e}_4}{e_3}{e_2}} \right\} T\left\{ {\left[ {{{\bar \psi }_1}({{\bar e}_1}{\psi _1}){\phi _1}} \right]  \left[ {{{\bar \psi }_2}{\psi _2}{\phi _2}} \right]} \right\}\\
 \left\{ {{{\bar e}_4}{e_3}{e_2}} \right\} T\left\{ {\left[ {{{\bar \psi }_1}{\psi _1}{\phi _1}} \right]  \left[ {{{\bar \psi }_2}({{\bar e}_1}{\psi _2}){\phi _2}} \right]} \right\}
 \end{array}
 \label{Eq.paire1}
\end{equation}
Since $H_1$ and $H_2$ are both ``new'' (i.e. not connected with any particles or fields), they are equivalent, thus the two choices in Eq. (\ref{Eq.paire1}) are equivalent. Knowing this, we can suppress the equivalent diagrams by demanding $\bar{e}_1$ to connect with $H_1$ only, and by multiplying a factor 2 to take into account the other choice.
The grouping for Eq. (\ref{Eq.explicitSNforH1}) under this additional restriction becomes
\begin{equation}
 \begin{array}{l}
 2 \times \frac{1}{{2!}}T\left\{ {\left[ {({e_2}{{\bar \psi }_1})({{\bar e}_1}{\psi _1})} \right] \left[ {({e_3}{{\bar \psi }_2})({{\bar e}_4}{\psi _2})({\phi _1}{\phi _2})} \right]} \right\}\\
 2 \times \frac{1}{{2!}}T\left\{ {\left[ {({e_3}{{\bar \psi }_1})({{\bar e}_1}{\psi _1})} \right] \left[ {({e_2}{{\bar \psi }_2})({{\bar e}_4}{\psi _2})({\phi _1}{\phi _2})} \right]} \right\}
 \end{array}
 \label{Eq.twoDiagforH1}
\end{equation}
From Eq. (\ref{Eq.twoDiagforH1}) we see, the equivalent diagrams disappear as expected. What's more, the factor $1/2!$ is canceled exactly by the multiplier 2. Indeed, this would happen for arbitrary $1/N!$ (only in the vacuum to vacuum case would this cancel be uncompleted, where a factor $1/N$ remains.).

The above example illustrates the first trick of our algorithm, namely

\textbf{Prescription 1:}
 \begin{itemize}
   \item In the procedure of an operator scanning over the fields in the T\{\} brace to pair with, when it encounter a ``new'' $H_i$, its scanning procedure is terminated after scanning over this $H_i$. The factor $1/N!$ in Eq. (\ref{Eq.abbrSN}) is discarded (or replaced by a $1/N$ in the vacuum to vacuum case) from the beginning.
 \end{itemize}
By this prescription, the equivalent Feynman diagrams due to vertex-relabeling are prohibited.

\begin{figure}[htbp]
\begin{center}
\subfigure[]{\includegraphics[width=0.15\textwidth]{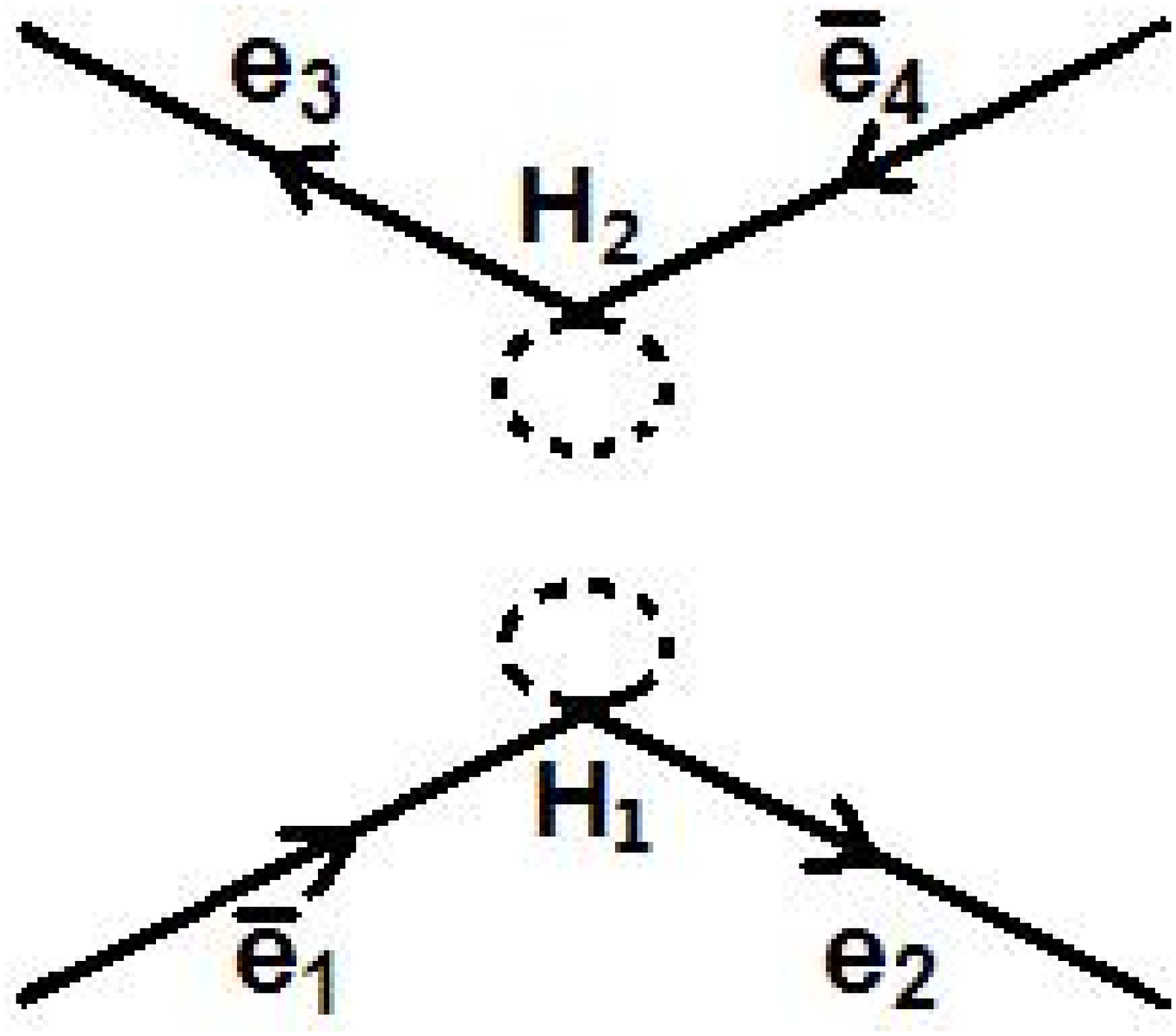}}
\subfigure[]{\includegraphics[width=0.15\textwidth]{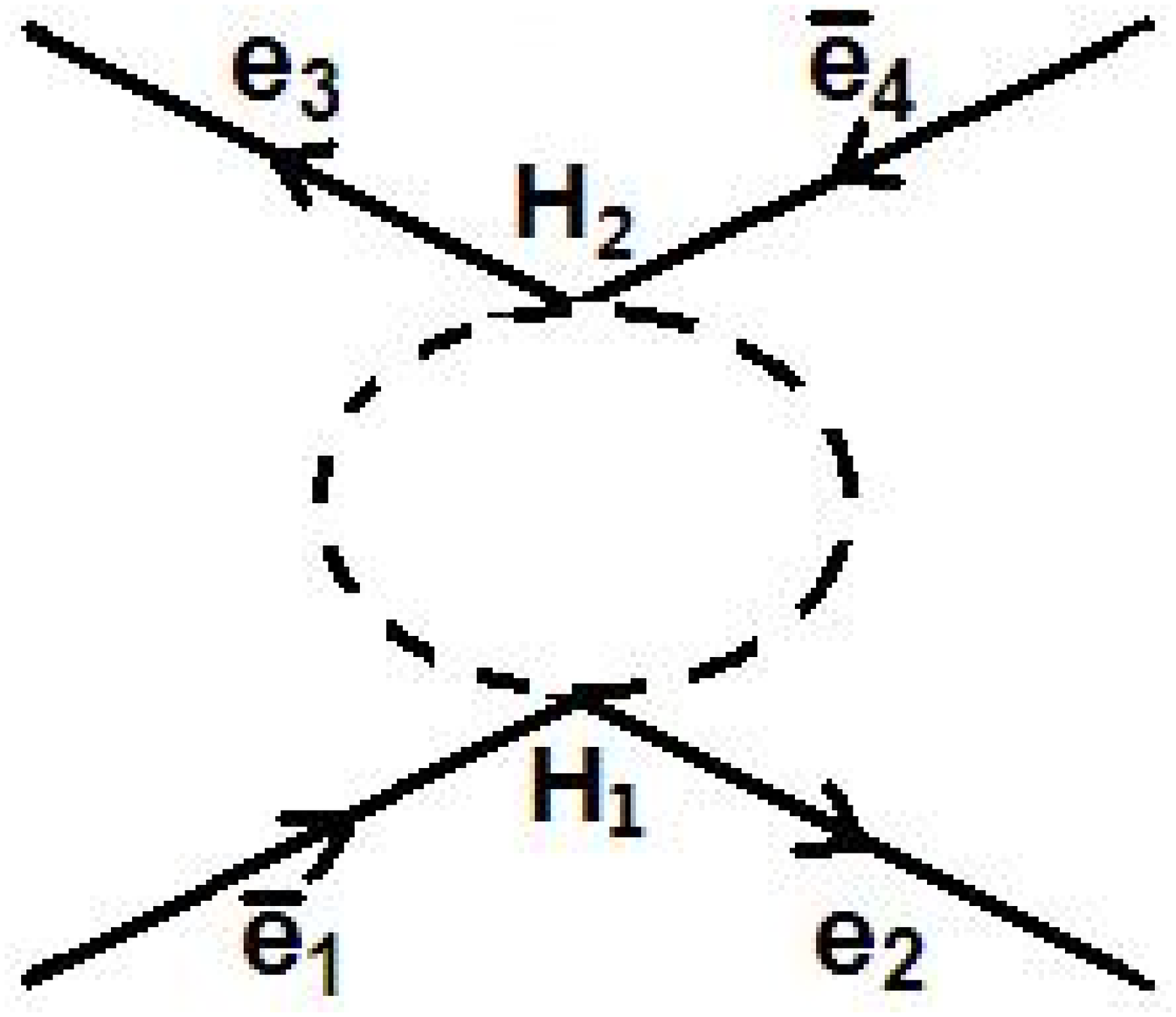}}
\subfigure[]{\includegraphics[width=0.15\textwidth]{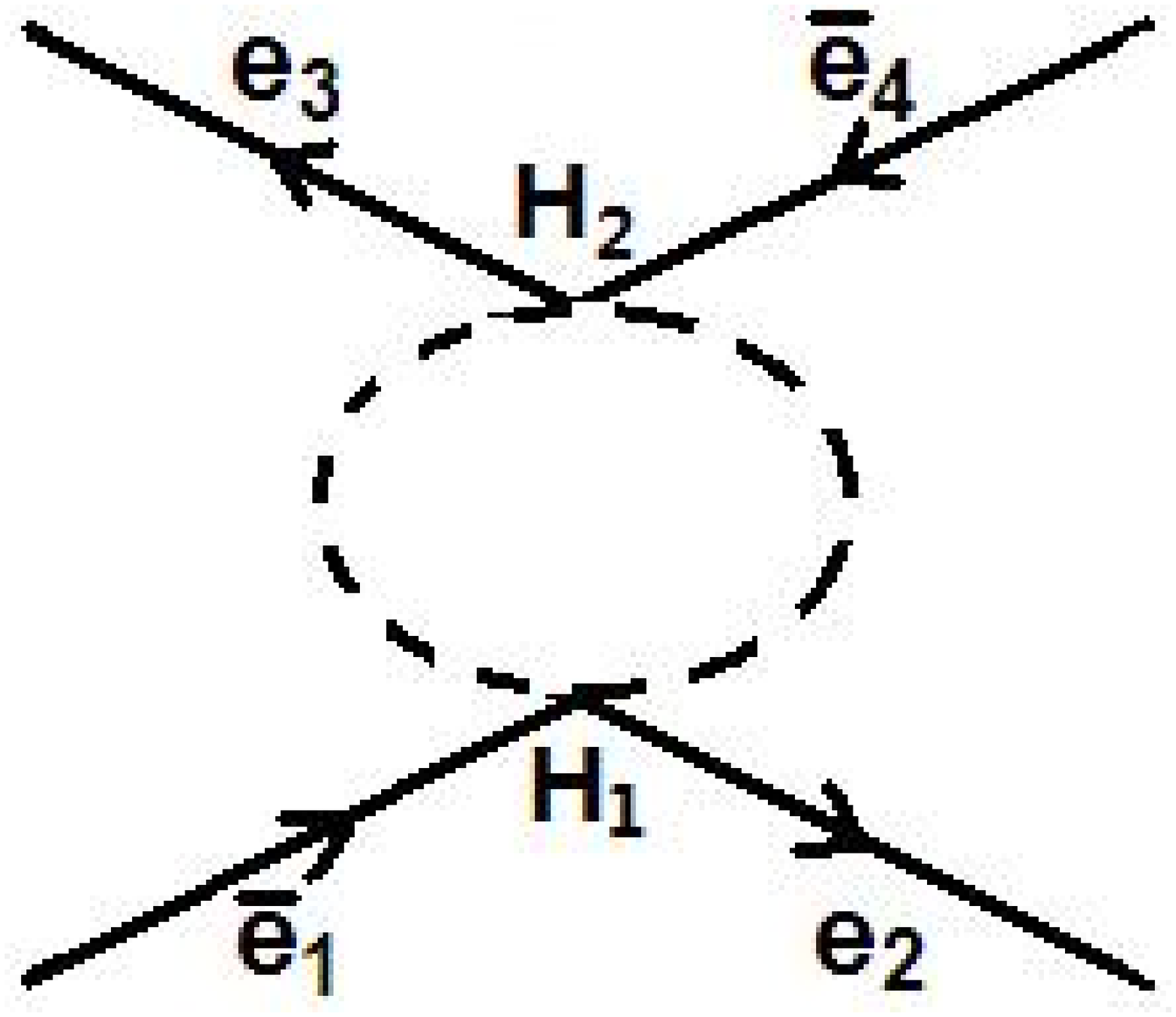}}
\subfigure[]{\includegraphics[width=0.15\textwidth]{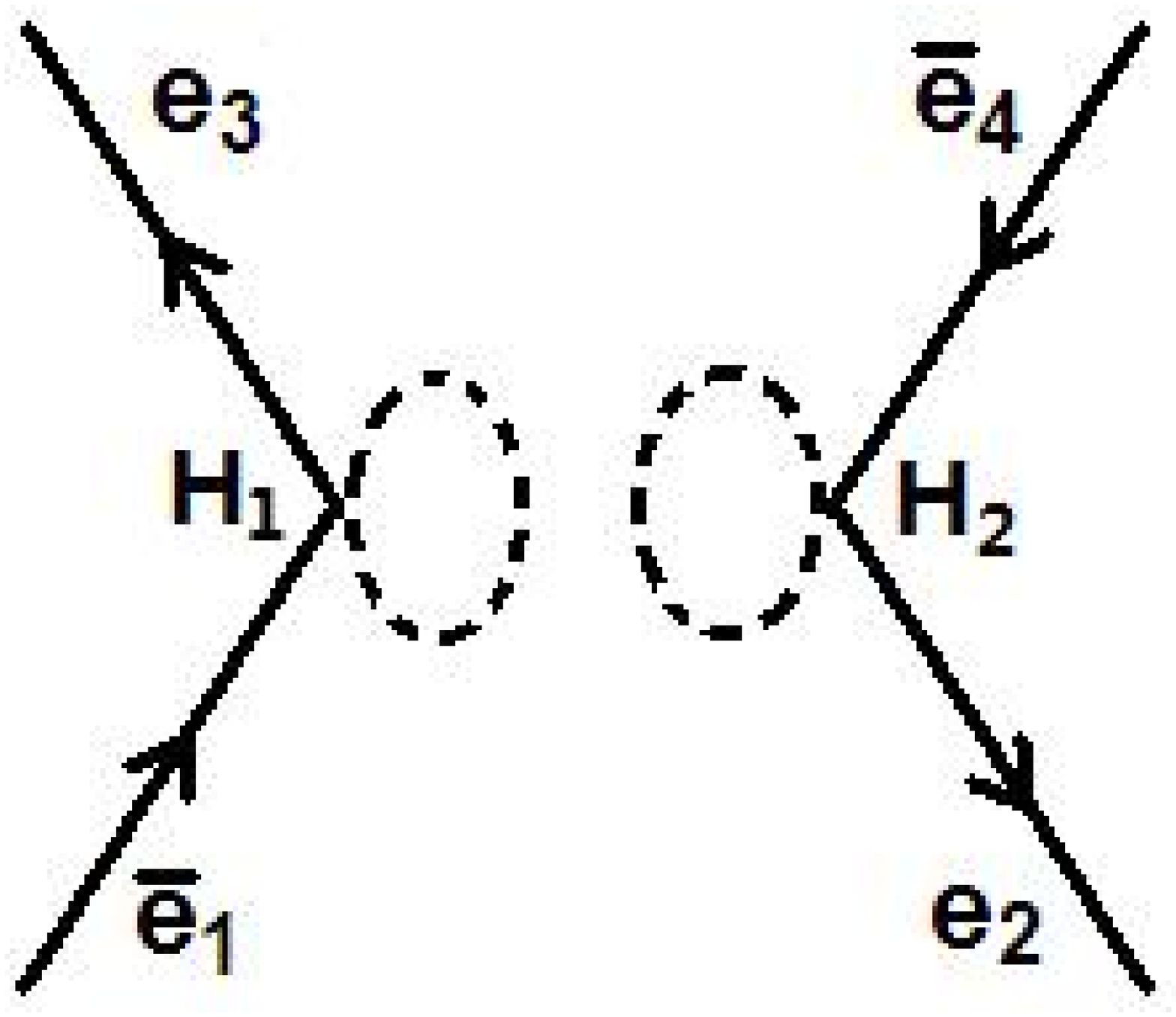}}
\subfigure[]{\includegraphics[width=0.15\textwidth]{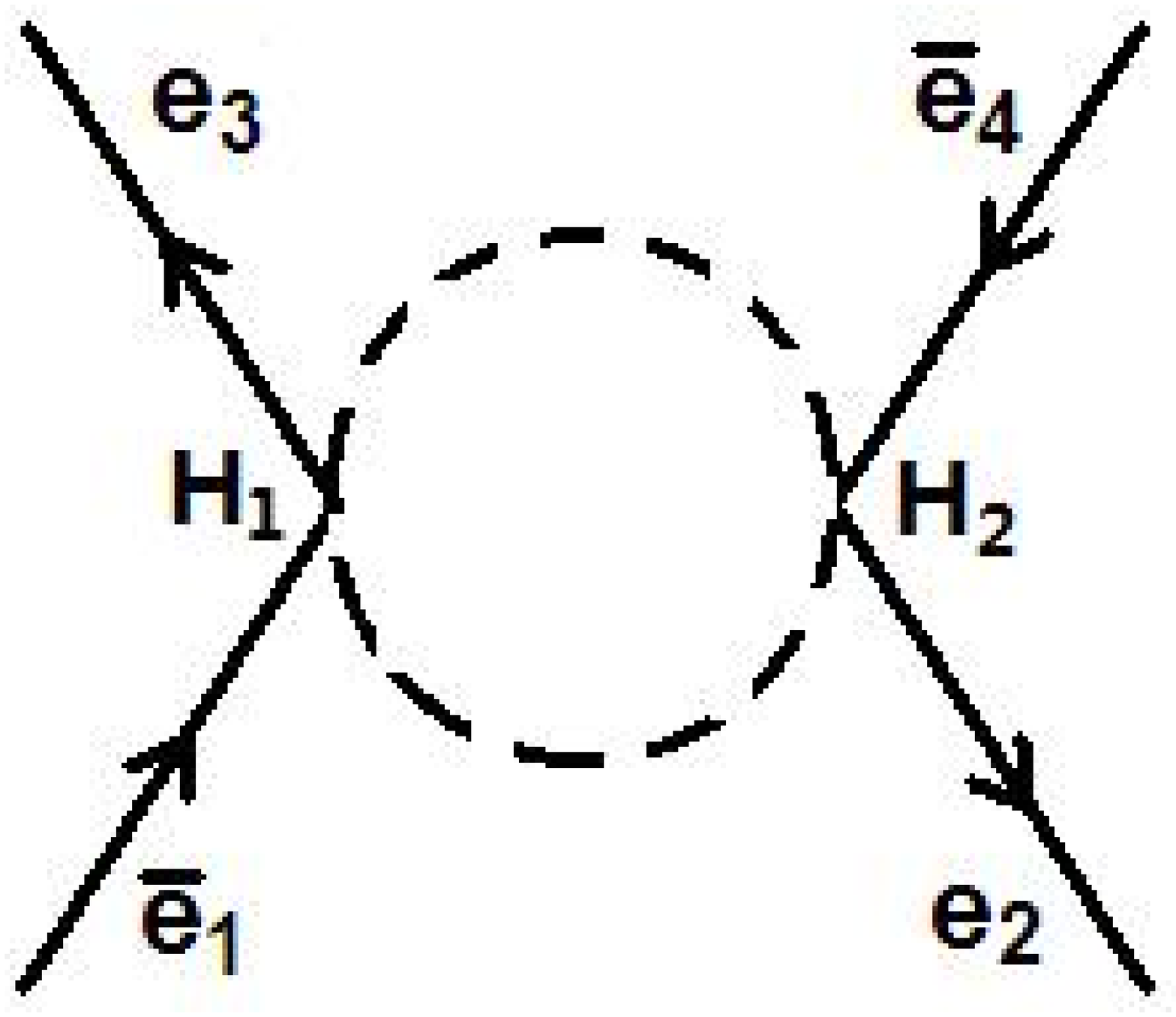}}
\subfigure[]{\includegraphics[width=0.15\textwidth]{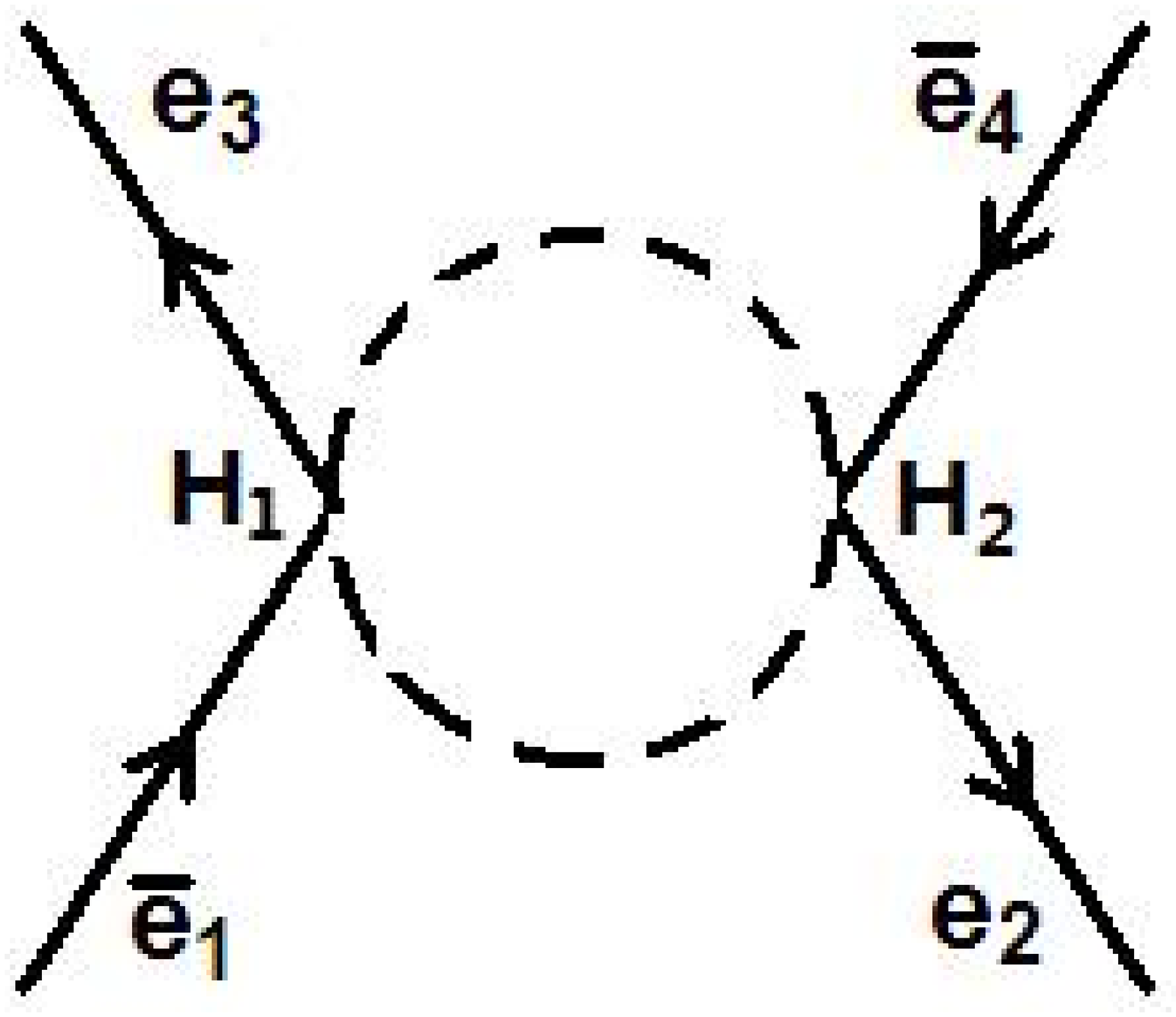}}
\end{center}
\caption{\label{SixDiagrams}six Feynman diagrams corresponding to the six ways of combinations in Eq. (\ref{Eq.sixDiag}). }
\end{figure}

Now, let's turn to another more complex example. We modify the previous example by change the physical model into
\begin{equation}
 H \propto {\bar \psi ^e}{\psi ^e}{\phi ^n}{\phi ^n}
 \label{Eq.H2}
\end{equation}
This leads to
\begin{equation}
 S_{1234}^2 = \frac{1}{{2!}}\left\{ {{{\bar e}_4}{e_3}{e_2}{{\bar e}_1}} \right\} \times T\left\{ {\left[ {{{\bar \psi }_1}{\psi _1}{\phi _1}{\phi _1}} \right] \left[ {{{\bar \psi }_2}{\psi _2}{\phi _2}{\phi _2}} \right]} \right\}
 \label{Eq.S2forH2}
\end{equation}
Following the ordered pairing procedure and the prescription 1 above, we derive from Eq. (\ref{Eq.S2forH2}) all the FD's as shown in Eq. (\ref{Eq.sixDiag}) and Figure \ref{SixDiagrams}.
\begin{equation}
 \begin{array}{l}
 T\left\{ {\left[ {({e_2}{{\bar \psi }_1})({{\bar e}_1}{\psi _1})({\phi _{1a}}{\phi _{1b}})} \right]  \left[ {({e_3}{{\bar \psi }_2})({{\bar e}_4}{\psi _2})({\phi _{2a}}{\phi _{2b}})} \right]} \right\}\\
 T\left\{ {\left[ {({e_2}{{\bar \psi }_1})({{\bar e}_1}{\psi _1})} \right]  \left[ {({e_3}{{\bar \psi }_2})({{\bar e}_4}{\psi _2})({\phi _{1a}}{\phi _{2a}})({\phi _{1b}}{\phi _{2b}})} \right]} \right\}\\
 T\left\{ {\left[ {({e_2}{{\bar \psi }_1})({{\bar e}_1}{\psi _1})} \right]  \left[ {({e_3}{{\bar \psi }_2})({{\bar e}_4}{\psi _2})({\phi _{1b}}{\phi _{2a}})({\phi _{1a}}{\phi _{2b}})} \right]} \right\}\\
 T\left\{ {\left[ {({e_3}{{\bar \psi }_1})({{\bar e}_1}{\psi _1})({\phi _{1a}}{\phi _{1b}})} \right]  \left[ {({e_2}{{\bar \psi }_2})({{\bar e}_4}{\psi _2})({\phi _{2a}}{\phi _{2b}})} \right]} \right\}\\
 T\left\{ {\left[ {({e_3}{{\bar \psi }_1})({{\bar e}_1}{\psi _1})} \right]  \left[ {({e_2}{{\bar \psi }_2})({{\bar e}_4}{\psi _2})({\phi _{1a}}{\phi _{2a}})({\phi _{1b}}{\phi _{2b}})} \right]} \right\}\\
 T\left\{ {\left[ {({e_3}{{\bar \psi }_1})({{\bar e}_1}{\psi _1})} \right]  \left[ {({e_2}{{\bar \psi }_2})({{\bar e}_4}{\psi _2})({\phi _{1b}}{\phi _{2a}})({\phi _{1a}}{\phi _{2b}})} \right]} \right\}
 \end{array}
 \label{Eq.sixDiag}
\end{equation}
We have add addition subscript $a$ or $b$ to the field $\phi$ in Eq. (\ref{Eq.sixDiag}) to distinguish
 different pairings.
Figure \ref{SixDiagrams}a and Figure \ref{SixDiagrams}b are two unconnected diagrams, which should be rejected in real calculations, we leave this task to the programming stage. Here we care about the equivalent diagrams (b) and (c) (or (e) and (f)). The arising of these equivalent diagrams can be traced back to the pairing of $\phi_{1a}$ with other fields. $\phi_{1a}$ faces three choices: $\phi_{1b}$, $\phi_{2a}$ and $\phi_{2b}$, where the paring with $\phi_{2a}$ and with $\phi_{2b}$ are equivalent. Knowing this, we can constraint $\phi_{1a}$ to pair with $\phi_{2a}$ only and represent the other equivalent choice by a multiplier 2. With this additional restriction, and discarding the unconnected ones, the grouping for Eq. (10) becomes
\begin{equation}
 \begin{array}{l}
 2 \times T\left\{ {\left[ {({e_2}{{\bar \psi }_1})({{\bar e}_1}{\psi _1})} \right]  \left[ {({e_3}{{\bar \psi }_2})({{\bar e}_4}{\psi _2})({\phi _{1a}}{\phi _{2a}})({\phi _{1b}}{\phi _{2b}})} \right]} \right\}\\
 2 \times T\left\{ {\left[ {({e_3}{{\bar \psi }_1})({{\bar e}_1}{\psi _1})} \right]  \left[ {({e_2}{{\bar \psi }_2})({{\bar e}_4}{\psi _2})({\phi _{1a}}{\phi _{2a}})({\phi _{1b}}{\phi _{2b}})} \right]} \right\}
 \end{array}.
 \label{Eq.diagOfPresc2}
\end{equation}

This example shows the second trick of our algorithm, that is

\textbf{Prescription 2:}
\begin{itemize}
  \item When an operator is paring with the fields in a vertex $H_i$, if there are $m$ identical unpaired fields in $H_i$ that can pair with this boson, the boson is restricted to pair with only one of them, and the other choices are represented by a multiplier $m$.
\end{itemize}

After the implementation of prescription 2, equivalent FD's originating from identical fields are greatly suppressed, and the symmetry factors are naturally generated. Note that the suppression by prescription 2 is incomplete, a small portion of equivalent FD's can still appear at two-loop order or higher. The correctness of the Feynman amplitude would not be hurt by the appearing of equivalent FD's, because
the symmetry factor attached to each FD would naturally compensate for it, e.g. a (4*FD) may appear as two equivalent FD's of (1*FD) and (3*FD). We don't try to utilize
a post eliminating procedure as done in many FD generators such as FeynArts, QGRAF and the grc part of GRACE to clean away the equivalent FD's, in order to keep the simplicity of the algorithm.

In effective field theory one may encounter the case of multiple identical \emph{fermions} appear in the interaction term, as
\begin{equation}
 H =  \cdots \psi  \cdots \psi  \cdots.
 \label{Eq.multFermion}
\end{equation}
In fact the two $\psi$'s in Eq. (13) are not surely identical because of their different positions in $H$: when they are paired by the same fermion, a difference of a factor $-1$ might arise. For simplicity, in our algorithm we just treat these fermions as {\it un-identical} ones (see the input file description in section \ref{sec.Cprogram} for more). The readers might wish to find a more elegant way to deal with this problem.

Minus signs coming from the interchange of fermions in the pairing procedure need also to be correctly produced, we leave this task to the programming stage.

Overall, the ``pairing order'' the ``prescription 1'' and the ``prescription 2'' form the main parts of our algorithm for FD generation.

\section{A example C program realizing the algorithm}\label{sec.Cprogram}

In this section, we briefly discuss a small C implementation of the algorithm proposed in section \ref{sec.algorithm}.

The C code can be download freely at the website \cite{codeWeb:2012}. In the C code, two independent functions, \texttt{judgeConnection()} and \texttt{fermionAcrossSign()}, are made to perform the tasks of judging the connectivity of a FD and the sign from fermion interchanging respectively.
 \texttt{judgeConnection()} is implemented after a FD is generated. \texttt{fermionAcrossSign()} is implemented in the FD making stage: every time an fermion operator pairs with another fermion field, the \texttt{fermionAcrossSign()} is called once. Other functions can be learned from the source code \cite{codeWeb:2012} directly.

The input file of the program is simple and clear. The following is an input file example corresponding to Eq. (10).

\scriptsize
\begin{spacing}{0.8} 
\begin{verbatim}
============================================================================
             The physical model (i.e. the relevent interactions)
instructions:
'U' means a Boson having no charge, i.e. it has no antiparticle, e.g. photon
'C' means a Boson which has antiparticle, e.g. W+ and W-
'F' means a Fermion
============================================================================
total number of interactions >>>
1
list of the interactions >>>
3> F1 F-1 U1:2

============================================================================
                          The physical process
============================================================================
total number of incoming and outgoing particles >>>
4
the incoming and outgoing particles >>>
2> F1 F-1
2> F1 F-1
total loops >>>
1
\end{verbatim}
\end{spacing}
\normalsize

In the model-defining part of the input file, the `\texttt{Ui}' or `\texttt{Ci}' can be followed by a `\texttt{:n}' which defines the multiplicity of the particle `\texttt{Ui}' or `\texttt{Ci}' in an interaction. If a user also want to write some interaction with multiple identical \emph{fermions}, she should write as ``...\texttt{Fi}...\texttt{Fi}...'', i.e. write each and every fermion explicitely.

The outputs corresponding to the above input file look like:

\scriptsize
\begin{spacing}{0.8} 
\begin{verbatim}
============================================================================
             The physical model (i.e. the relevent interactions)
============================================================================
>>> the interactions
I1:	(F1,F-1,U1:2)

============================================================================
                          The physical process
============================================================================
>>> the external particles
P1:	(F-1)
P2:	(F1)
P3:	(F1)
P4:	(F-1)
>>> the internal vertices
H1
H2
H3
H4

============================================================================
                          The Feynman diagrams
============================================================================
(1)
P1-->H1 (I1.leg[2])
P2-->H1 (I1.leg[1])
P3-->H2 (I1.leg[1])
P4-->H2 (I1.leg[2])
H1-->H2 (I1.leg[3]-->I1.leg[3])
H1-->H2 (I1.leg[3]-->I1.leg[3])
symmetryFactor = 2 *(1/2!)*(1/2!) = 1/2
fermionSign=-1
------------------------------------------------
(2)
P1-->H1 (I1.leg[2])
P2-->H2 (I1.leg[1])
P3-->H1 (I1.leg[1])
P4-->H2 (I1.leg[2])
H1-->H2 (I1.leg[3]-->I1.leg[3])
H1-->H2 (I1.leg[3]-->I1.leg[3])
symmetryFactor = 2 *(1/2!)*(1/2!) = 1/2
fermionSign=1
------------------------------------------------
\end{verbatim}
\end{spacing}
\normalsize

The C program is powerful, yet small. It is a totally general purpose FD generator, that is, it can receive arbitrary user defined models and arbitrary processes and any loops. Yet, benefit from the simplicity of the algorithm, the C program is small, i.e., $\sim 500$ lines of code as a whole, and $\sim 200$ lines as the core part for realizing the algorithm. It is also very fast: 3860 FD's of $u\bar{u}\rightarrow t\bar{t}$ at two loop order in QCD model can be generated in 0.015 seconds in a normal PC (compared to minutes by FeynArts).

However, as only an illustration of our algorithm, it still lacks many functionalities which one might find useful: showing the Feynman diagrams in a graph view mode, picking out the one-particle irreducible diagrams, to name a few.

\section{Checking the correctness of the algorithm/C-program\label{sec.checkment}}

In order to check the correctness of the proposed algorithm and the C-program, we use FeynArts and our C-program to generate FD's for the process $u\bar u\rightarrow t\bar t$ in three different models
\[
\begin{array}{l}
  \text{MODEL1:  }  u\bar u g + t\bar t g \\
  \text{MODEL2:  }  u\bar u g + t\bar t g + g^3 \\
  \text{MODEL3:  }  u\bar u g + t\bar t g + g^4
\end{array}
\]
and at three different loop orders: tree, one-loop and two-loop.
Table 1 shows the total number of FD's generated by the two programs, from which one see, the two programs agree with each other except for the MODEL2 and MODEL3 at two-loop order. These differences are attributed to the fact that our algorithm generates more equivalent FD's in these cases. In table 2 and table 3 we show explicitly all the equivalent FD's generated by our C-program and by FeynArts (When doing the comparison, we found that FeynArts also generate some equivalent FD's). Combining Table \ref{totalFDNum}, Table \ref{EquivFDsM2} and table \ref{EquivFDsM3}, one see, the two programs agree exactly with each other.

\begin{table}[htb]
\caption{\label{totalFDNum}Comparison between the total numbers of FD's generated by our C-program and FeynArts.
The numbers at the left (right) of the comma are from our C-program (FeynArts).}
\begin{tabular}{|c|c|c|c|}
\hline\hline
\begin{tabular}{c}
\\
\hline
tree\\
one-loop\\
two-loop\\
\end{tabular}
 &
\begin{tabular}{c}
MODEL1\\
\hline
1, 1\\
18, 18\\
303, 303\\
\end{tabular}
 &
\begin{tabular}{c}
MODEL2\\
\hline
1, 1\\
28, 28\\
825, 764\\
\end{tabular}
 &
\begin{tabular}{c}
MODEL3\\
\hline
1, 1\\
19, 19\\
369, 355\\
\end{tabular}
\\
\hline\hline
\end{tabular}
\end{table}

\begin{table}[htb]
\caption{\label{EquivFDsM2}Equivalent FD's generated by our C-program (first row) and FeynArts (second row) under the model MODEL2. The numbers in bracket are the ID number of the FD's generated in our C-program.}
\begin{tabular}{p{460pt}}
\hline\hline
(12)(15), (13)(16), (14)(17), (19)(20), (32)(35), (33)(36), (34)(37), (39)(40), (51)(53), (52)(60), (56)(61), (64)(67), (65)(66), (68)(70), (69)(72), (73)(74), (75)(78), (76)(77), (79)(81), (82)(97), (83)(98), (84)(99), (85)(100), (86)(102), (87)(103), (88)(104), (89)(105), (90)(107), (91)(108), (92)(109), (93)(110), (94)(111), (95)(112), (96)(113), (101)(121), (106)(122), (114)(117), (115)(118), (116)(119), (123)(126), (124)(125), (137)(139), (138)(146), (142)(147), (188)(190), (189)(197), (193)(198), (329)(332), (330)(333), (331)(334), (339)(340), (342)(345), (343)(346), (344)(347), (349)(350), (375)(376), (387)(389), (388)(396), (392)(397), (434)(436), (435)(443), (439)(444),  (475)(478), (476)(479), (477)(480), (483)(484) \ \ \ \ \ \ \ \ \ \ \ \ \ \ \ \ \ \ \ \ \ \ \ \ \ \ \ \ \ \ \ \ \ \ in total: 66 pairs
 \\
\hline
\includegraphics[width=1.0in]{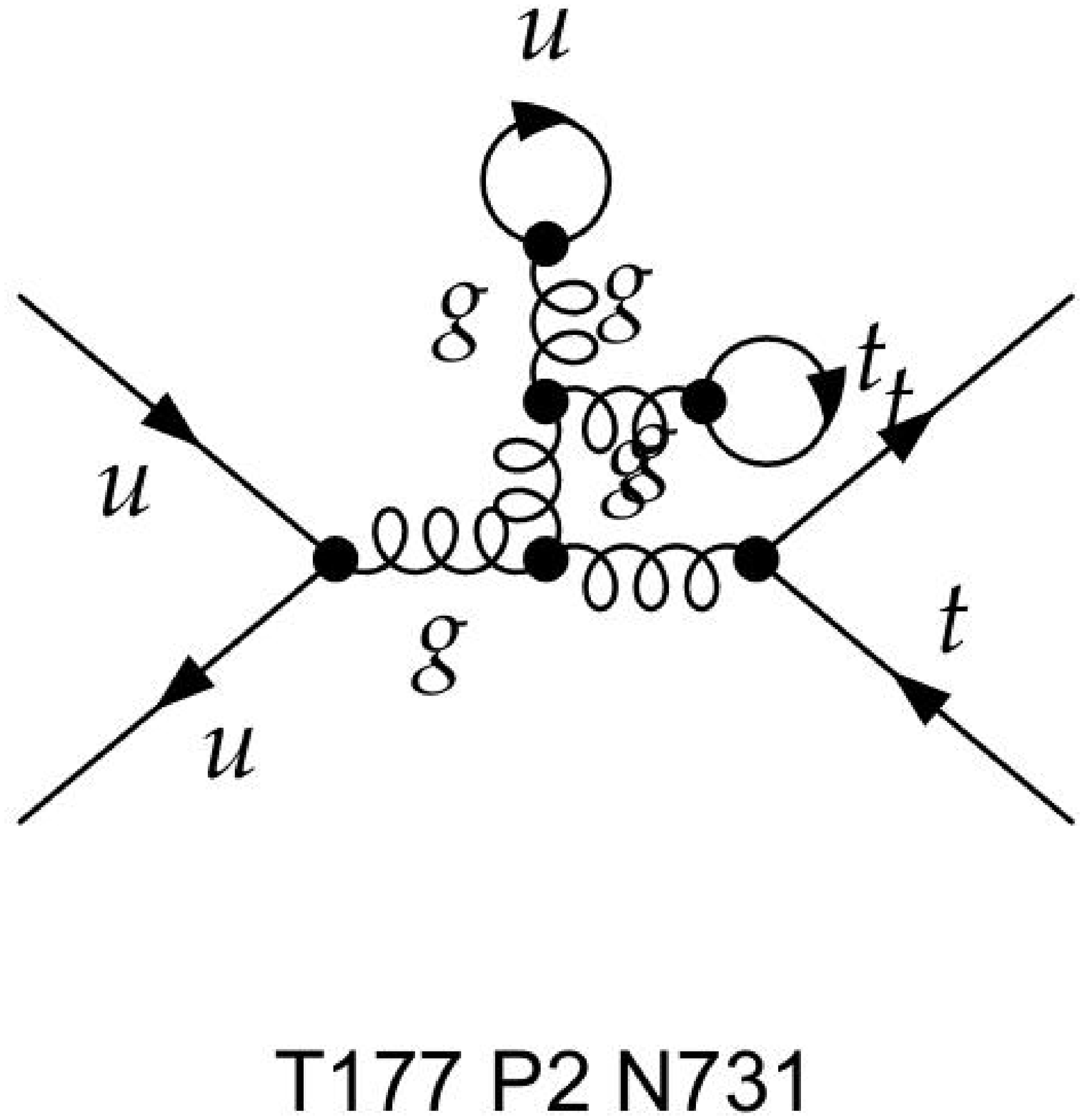}
\includegraphics[width=1.0in]{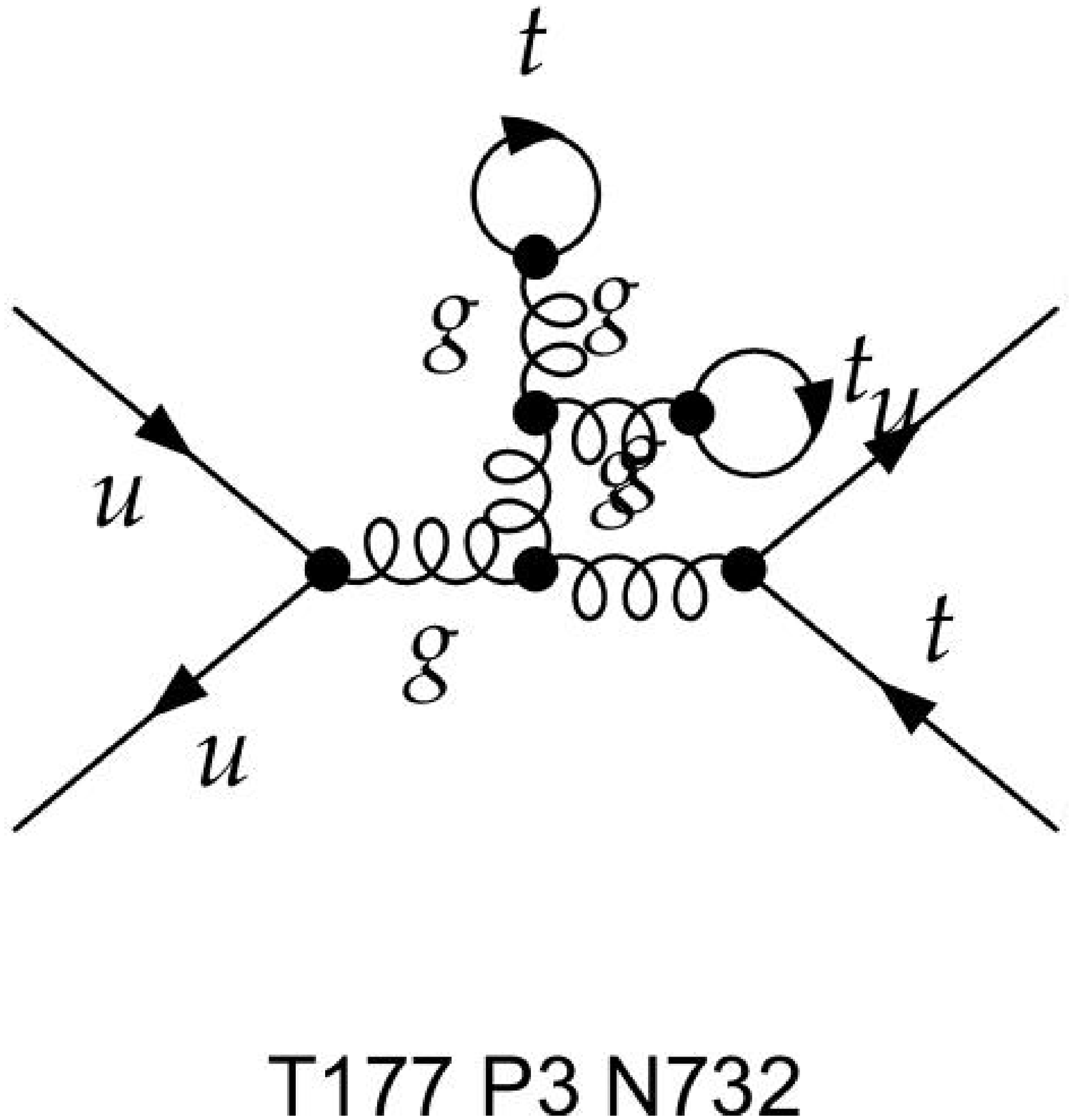}
\includegraphics[width=1.0in]{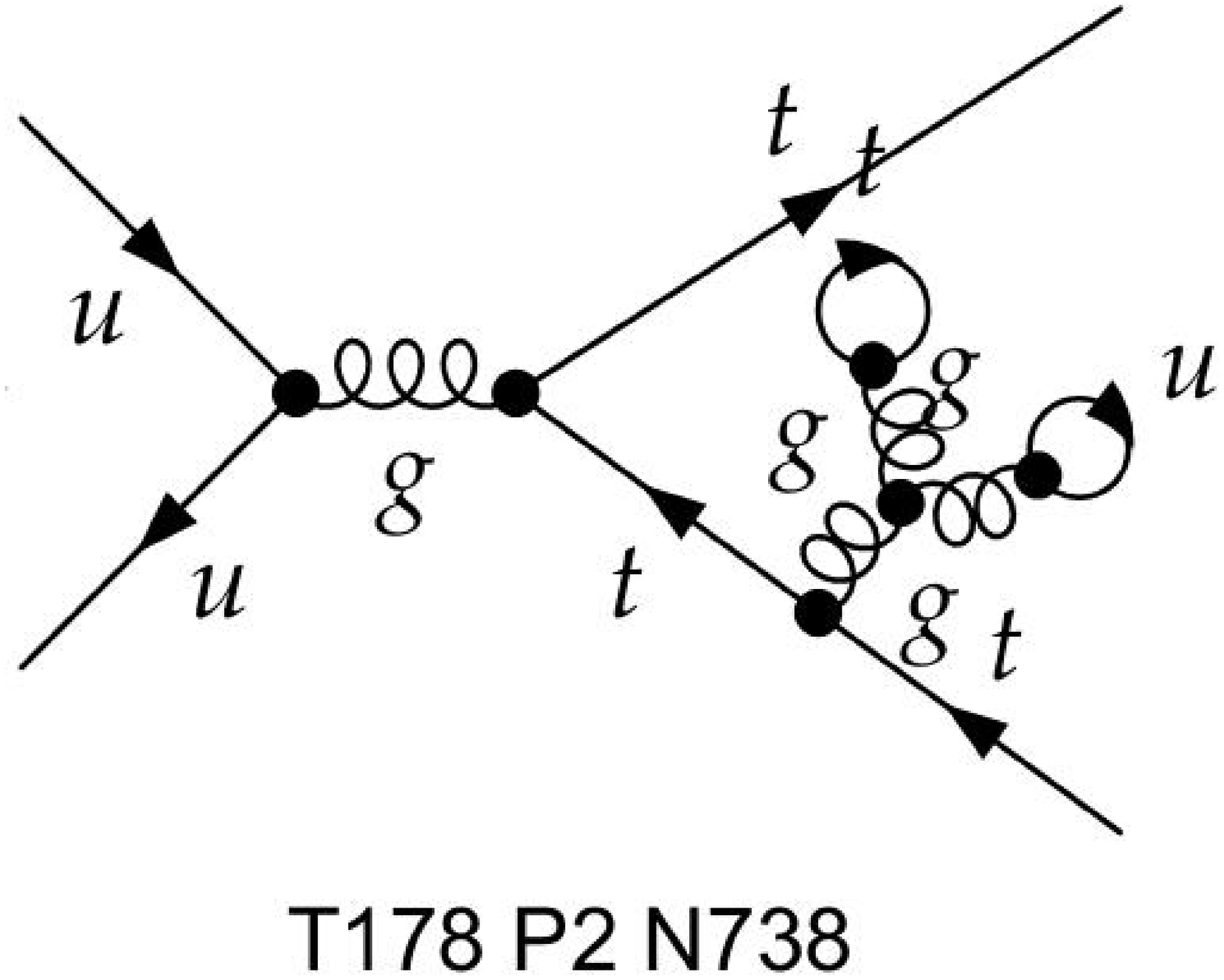}
\includegraphics[width=1.0in]{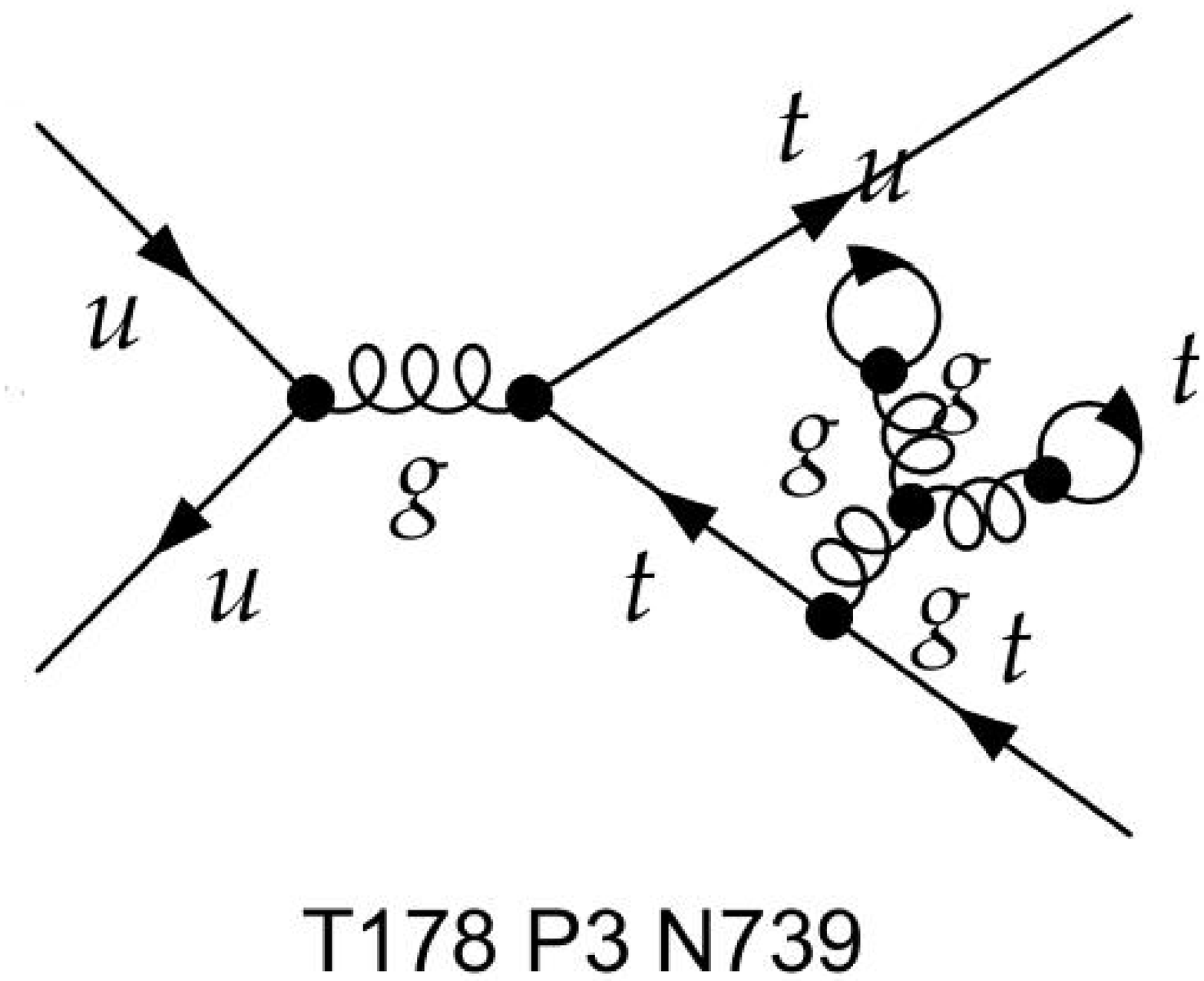}
\includegraphics[width=1.0in]{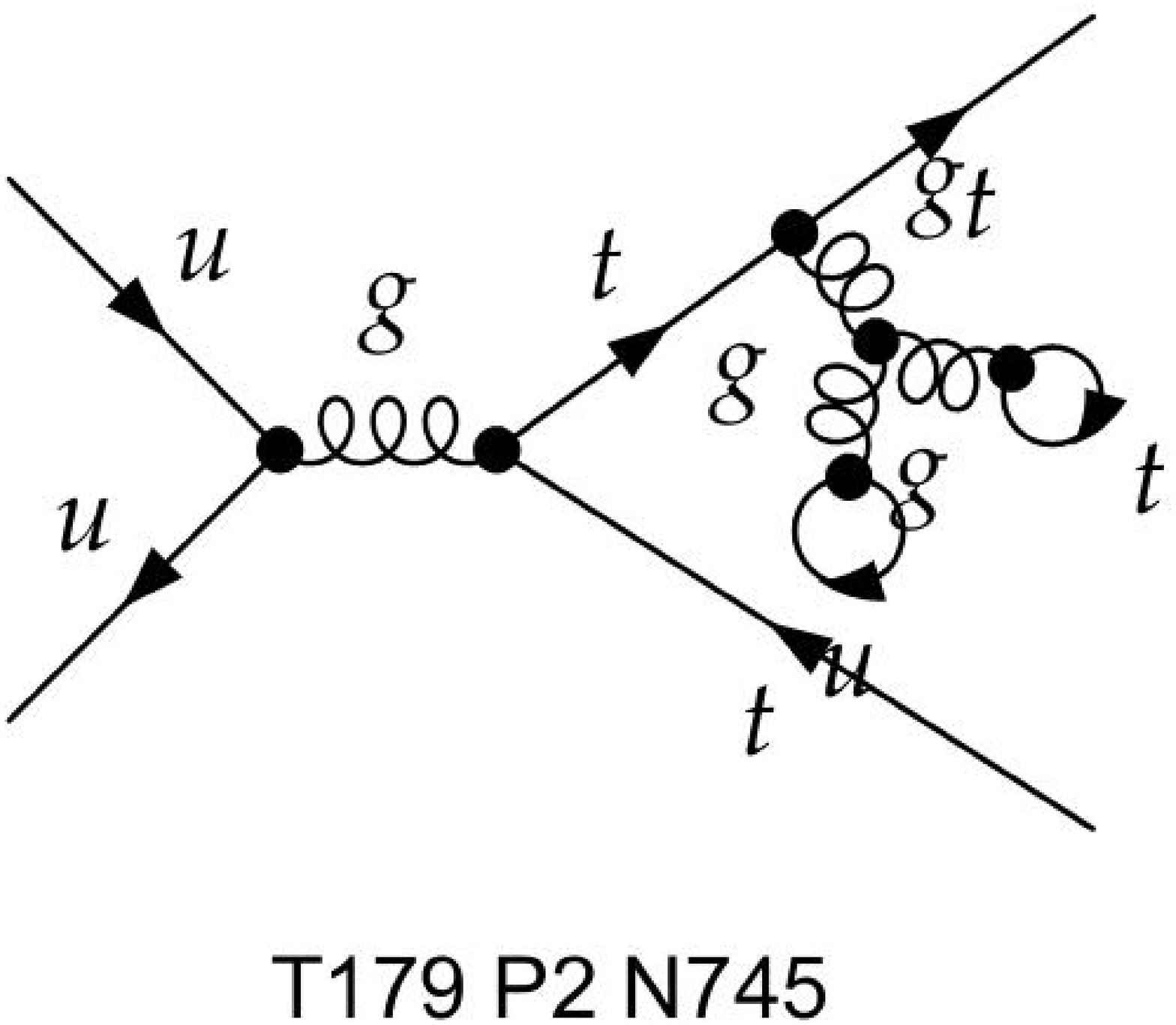}
\includegraphics[width=1.0in]{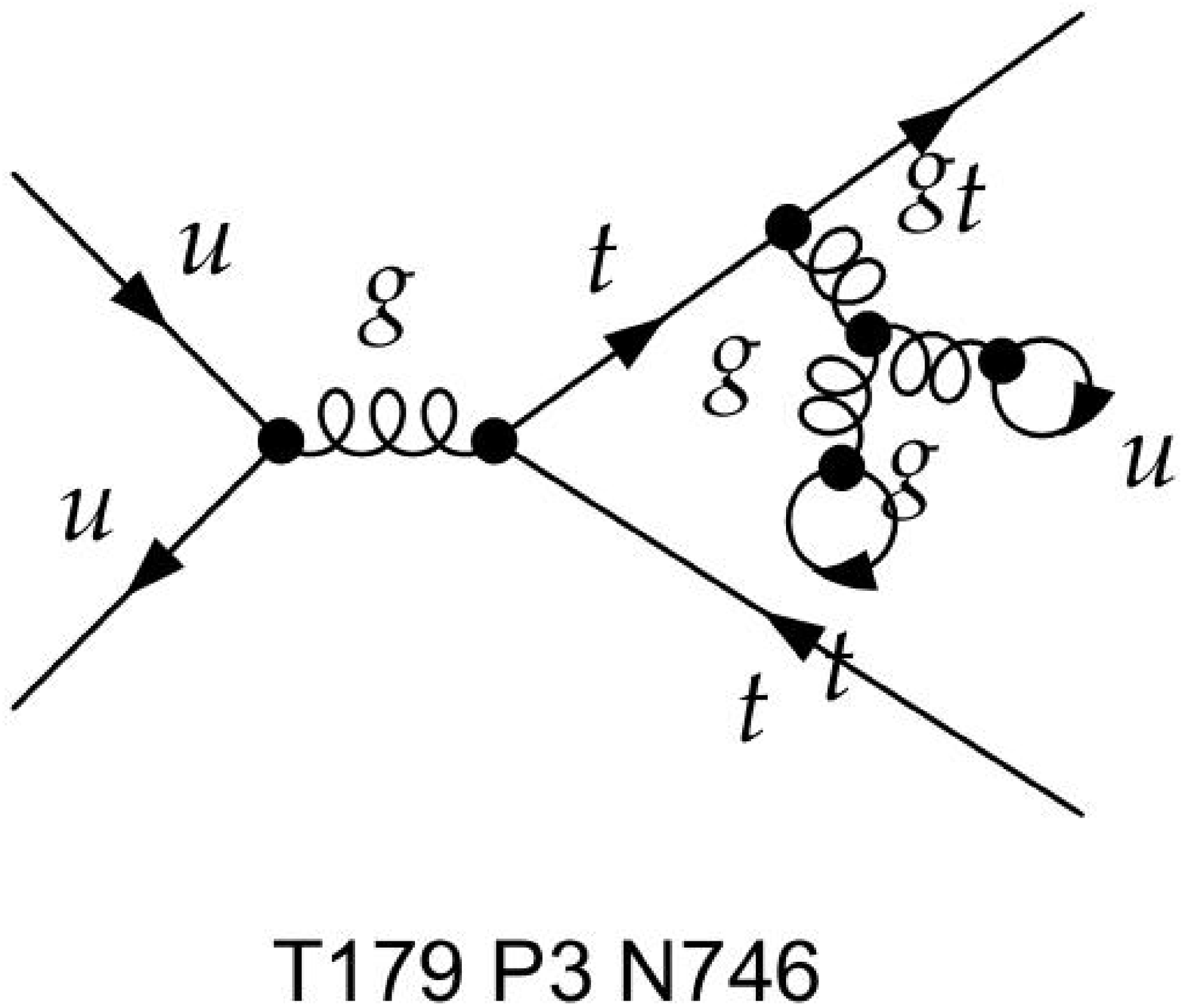}
\includegraphics[width=1.0in]{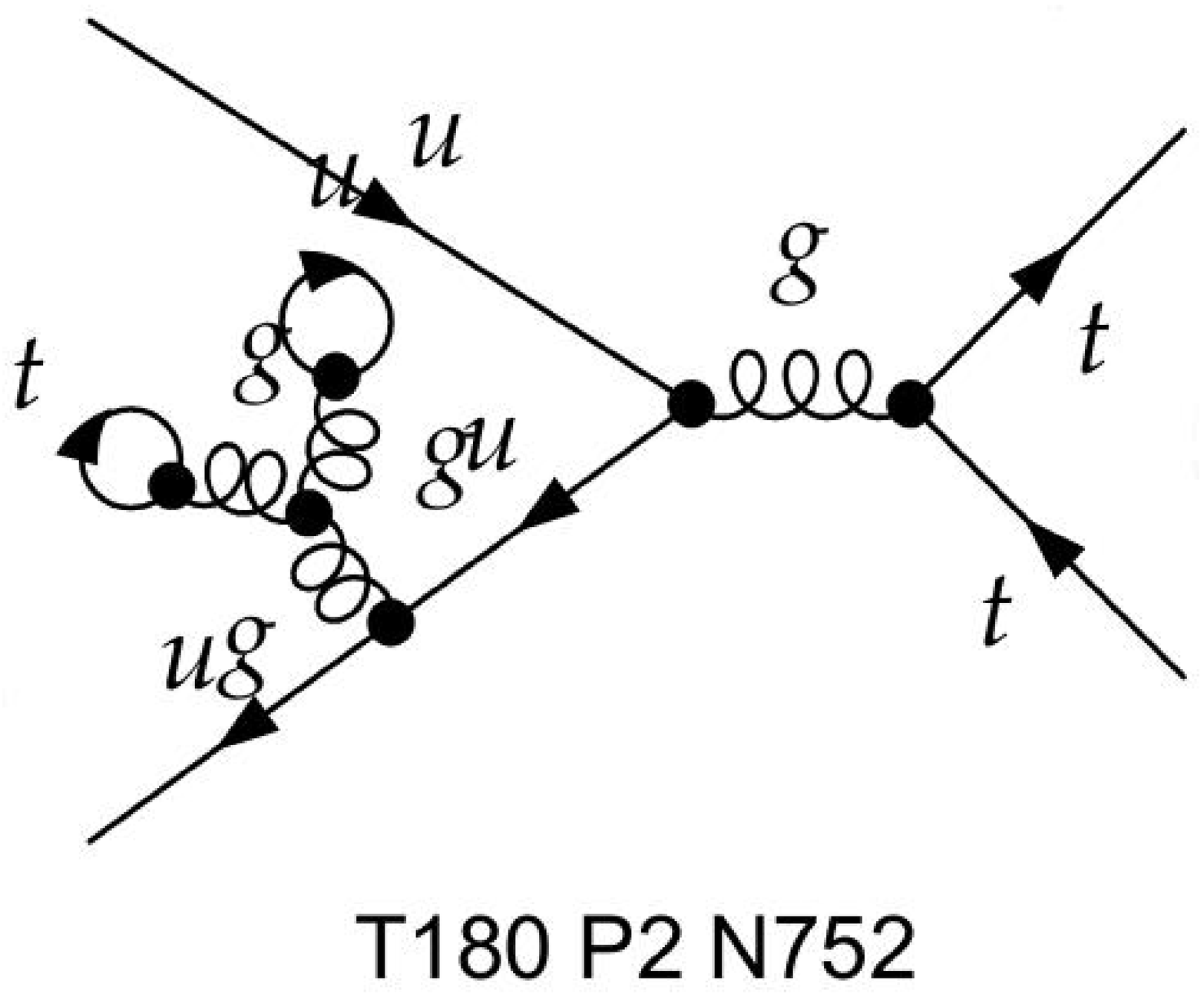}
\includegraphics[width=1.0in]{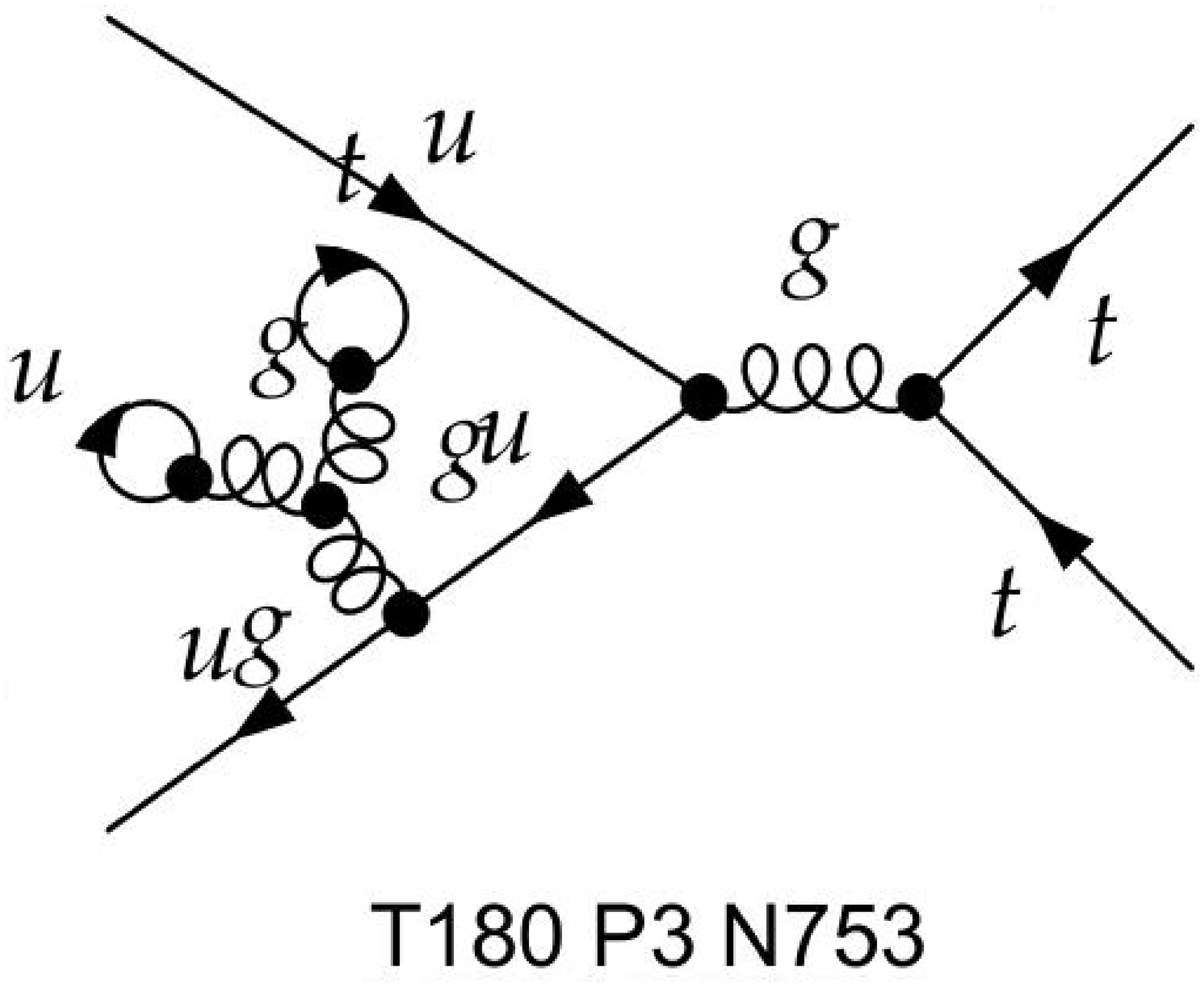}
\includegraphics[width=1.0in]{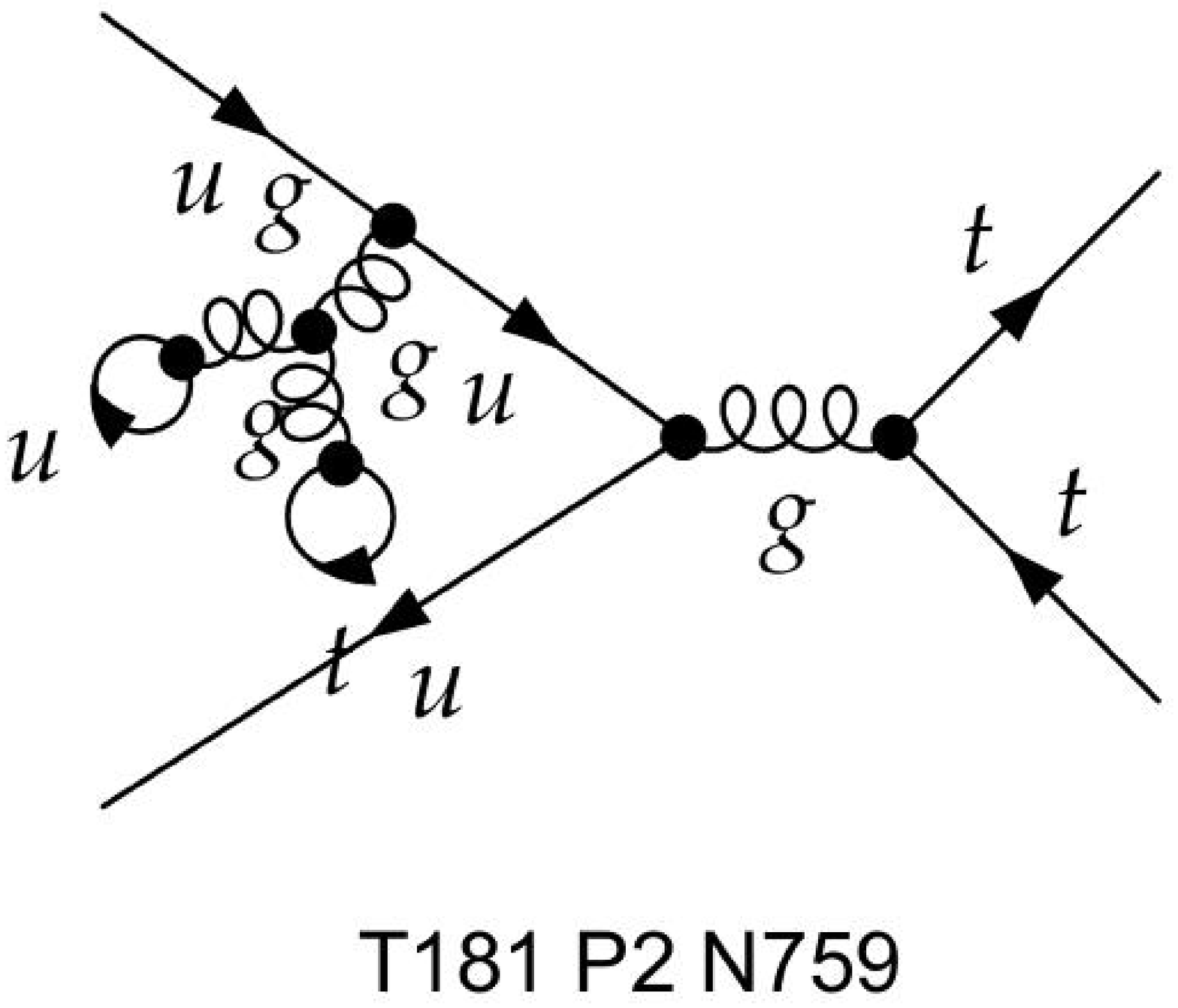}
\includegraphics[width=1.0in]{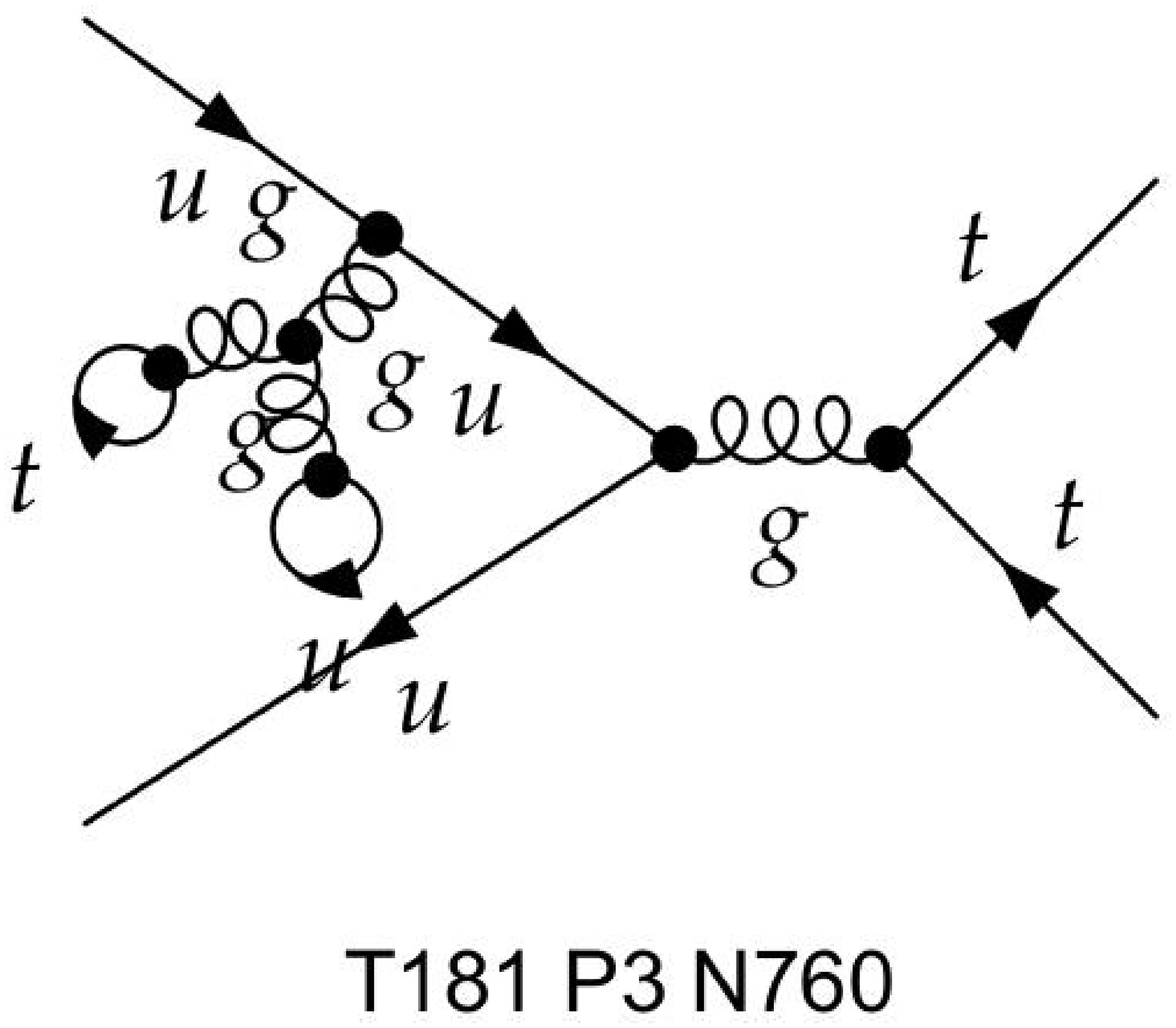}
 \ \ \ \ \ \ \ \ \ \ \ \ \ \ \ \ \ \ \ \ in total: 5 pairs
 \\
\hline\hline
\end{tabular}
\end{table}

\begin{table}[htb]
\caption{\label{EquivFDsM3}Same as Table \ref{EquivFDsM2} but with the model changed to MODEL3.}
\begin{tabular}{p{460pt}}
\hline\hline
(10)(11), (21)(22), (25)(26), (30)(31), (32)(33), (34)(35), (43)(45), (44)(46), (69)(71), (70)(72), (141)(142), (152)(154), (153)(155), (176)(178), (177)(179) \ \ \ in total: 15 pairs
 \\
\hline
\includegraphics[width=1.0in]{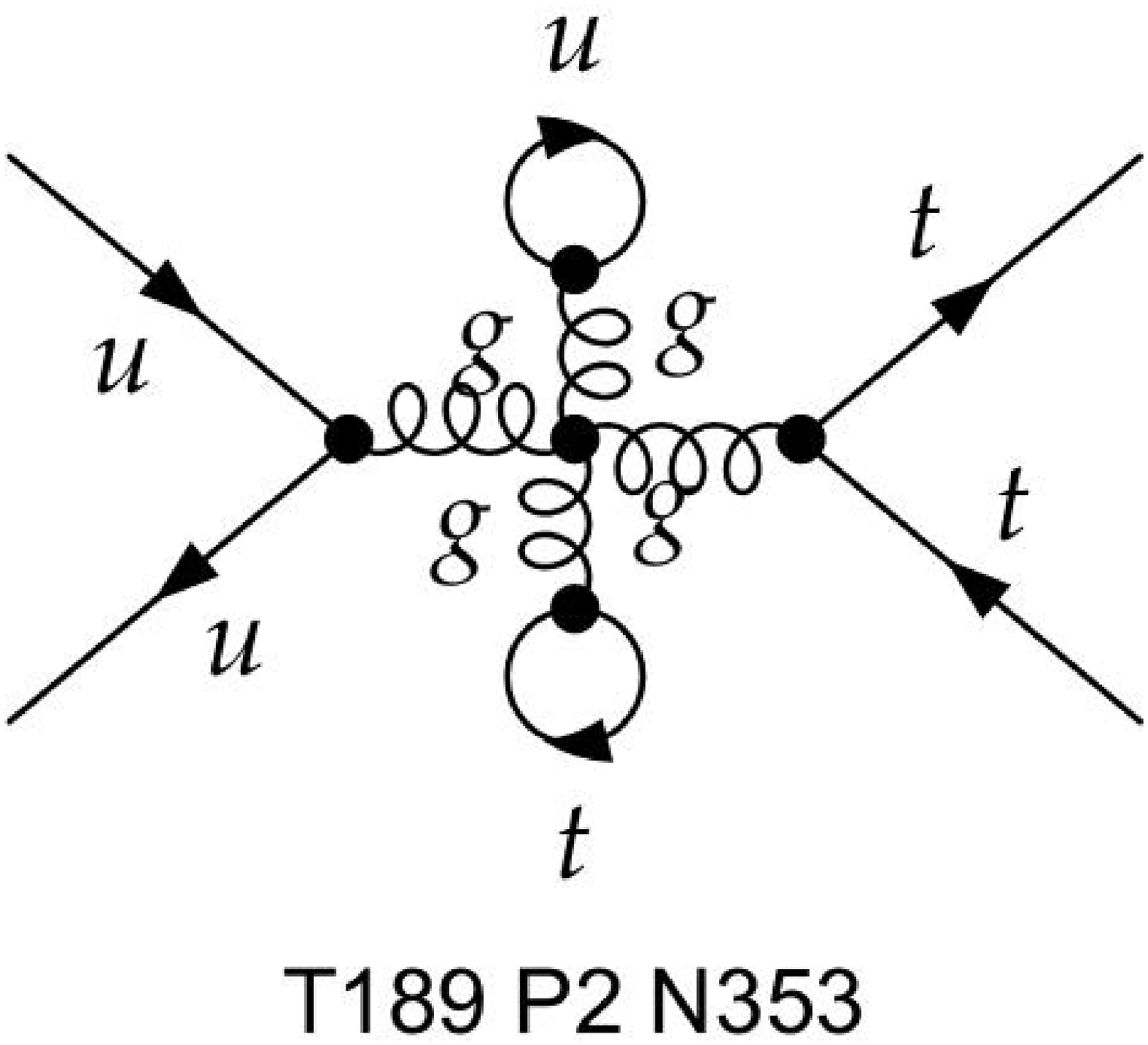}
\includegraphics[width=1.0in]{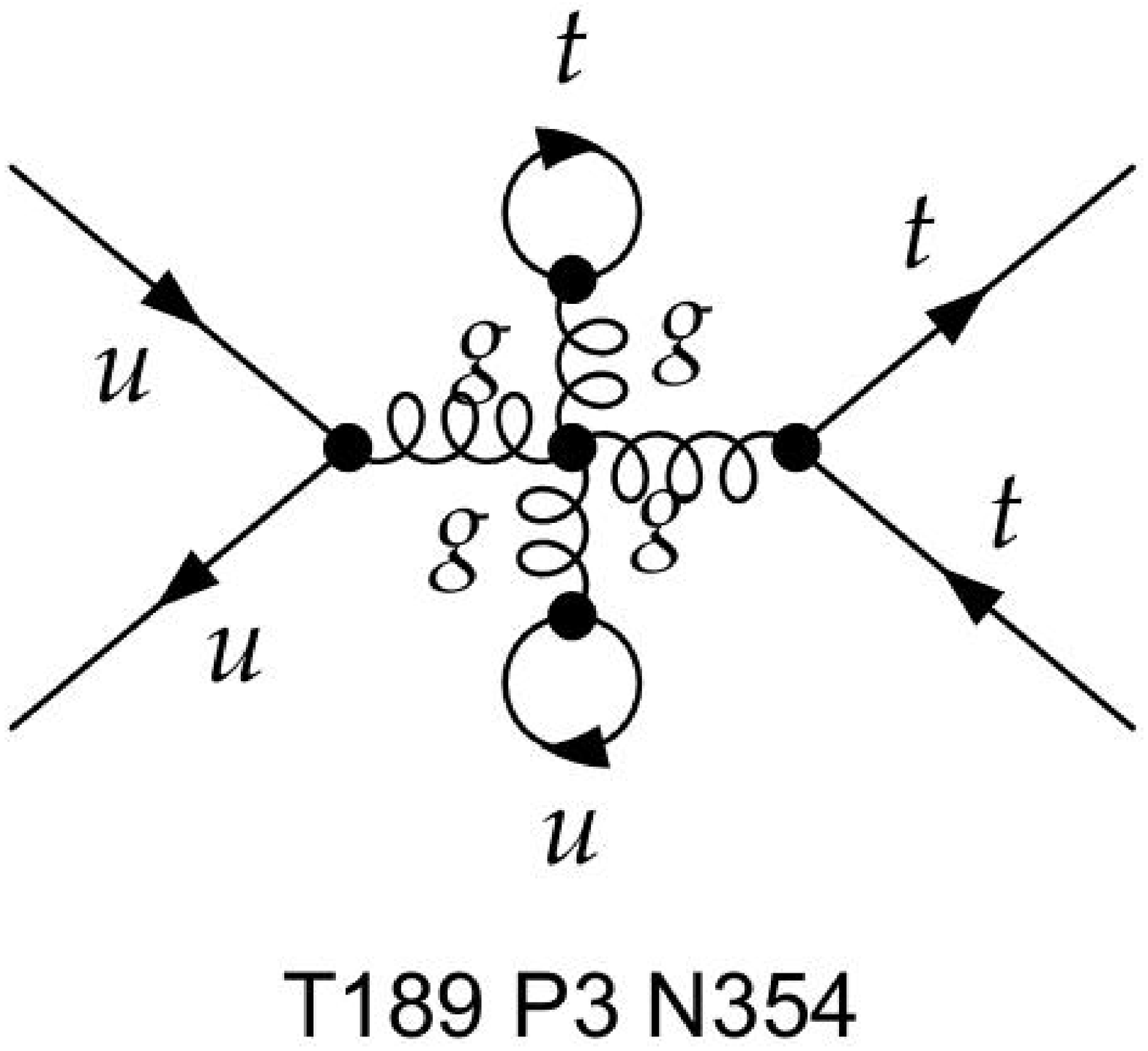}
 \ \ \ \ \ \ \ \ \ \ \ \ \ \ \ \ \ \ \ \ \ \ \ \ \ \ \ \ \ \ \ \ \ \ \ \ \ \ \ \ in total: 1 pairs
 \\
\hline\hline
\end{tabular}
\end{table}

\section{Conclusion and Discussion\label{sec.conclusion}}

An algorithm for FD generation is proposed in this paper. It is simple in concept, easy for coding. A C-program realizing this algorithm is presented. It is small in size and run fast, yet is a totally general FD generator: it receives arbitrary user defined model and arbitrary process as input and generates FD's at any order.

We hope this simple algorithm could make life easier for high energy researchers, who would like to make their own FD generators and add functionalities to their own taste. Researchers who are developing new calculation techniques may find the algorithm of special interests, as it could help in both testing new ideas and constructing new calculation tools. It can also be useful for researchers working on the effective field theory due to its `arbitrary user defined model' property.

{\em Acknowledgements:} This work is supported in part by the Developing Foundation of CAEP (No. 2009A0203013).


\end{document}